\documentclass[11pt]{article}
\usepackage[margin=1in]{geometry}
\usepackage{amsmath,amssymb}
\usepackage{bm}
\usepackage{graphicx}
\usepackage{booktabs}
\usepackage{siunitx}
\usepackage{tikz}
\usetikzlibrary{arrows.meta}
\usepackage{subcaption}
\usepackage[numbers,sort&compress]{natbib}
\usepackage{authblk}
\usepackage{float}
\usepackage{placeins}
\usepackage[hidelinks]{hyperref}
\graphicspath{{figs/}}

\title{\textbf{A Two-Dimensional Physics-Informed Neural Network Model of a Hall Effect Thruster Plasma: Self-Consistent Electrostatics and Physical-Consistency Diagnostics}}
\author{Josh Adjedj
}
\author{Subrata Roy\thanks{Corresponding Author. Email: roy@ufl.edu}}
\affil{Applied Physics Research Group, Department of Mechanical and Aerospace Engineering, University of Florida, Gainesville, Florida 32611, USA}

\begin{document}
\maketitle

\noindent\rule{\textwidth}{0.4pt}
\begin{center}\textbf{Abstract}\end{center}
\noindent
We present a two-dimensional axisymmetric physics-informed neural network (PINN) of the discharge channel of the 5-kW-class P5 Hall thruster at \SI{300}{\volt} and \SI{5.4}{\ampere}, which solves eleven coupled plasma fields and assimilates 145 internal probe measurements per field reconstructed from the data of Haas and Gallimore. The electric potential is obtained from Poisson's equation in charge-separation form, $n_e = n_i(1+\eta)$, with the charge separation $\eta$ bounded by construction, which keeps the stiff electrostatic constraint ($\alpha_p \approx 7.24\times10^{7}$) active and trainable. The reference solution agrees with the reconstructed profiles on which it is trained to relative $L^2$ errors of $0.017$, $0.073$ and $0.205$ for the plasma potential, electron temperature and ion density, and held-out tests show the potential to be far more robust than the ion density to the withdrawal of measurement lines, the density reconstruction depending on where the measurements are placed. We then test the solution against physical-consistency requirements that the data fit does not reveal: the global particle and current balances, the electron current implied by the momentum equation, the field-aligned electron force balance and the axial ion momentum budget. These diagnostics expose a neutral inflow six times the measured propellant flow, a bound in the potential representation that makes part of the measured potential unreachable, and a momentum-implied electron current, more than 99\% field-aligned, that exceeds the discharge current by orders of magnitude. Correcting the inflow and the anode potential, introducing a divergence-free magnetic field fitted to measured vacuum-field data, enlarging the potential representation, weighting the data by their measurement uncertainties, adding the electron pressure gradient, and imposing the field-aligned balance and the discharge current progressively yields a solution that conserves particles to within 3\%, holds the network current between 5.1 and \SI{5.4}{\ampere} along the channel, and places 97\% of the potential and 88\% of the temperature measurements within their uncertainties. Within the tested single-velocity fluid formulation, however, the measured ion density, potential and ion velocity could not be matched together: with the current imposed, the exit ion energy exceeds the available potential drop and is supplied by the ion momentum residual. The analysis identifies the field-aligned electron balance as the remaining obstacle to electron self-consistency.

\newpage

\section{Introduction}
In an era where human beings seek to establish a continuous presence on the Moon, Mars, and beyond, strengthening our knowledge and mastery of the different kinds of propulsion technologies is critical to the success of deep-space travel. Among electric propulsion technologies, the Hall effect thruster (HET) is the most popular. Developed from the concept of closed-drift accelerators and matured over five decades of flight history, HETs deliver specific impulses between roughly 1000 and \SI{3000}{\second} at powers from \SI{100}{\watt} to tens of kilowatts \citep{goebel2008,boeuf2017}, a propellant economy that chemical propulsion cannot reach. Originally limited to station-keeping and orbit raising of commercial satellites,the technology has progressed toward primary propulsion for deep-space missions: the 12.5-kW-class Advanced Electric Propulsion System (AEPS) Hall thruster, among the highest-power electric propulsion system yet developed for flight, has completed qualification testing \citep{watts2025}. This growing interest in electric propulsion is precisely why a deeper, faster, and more flexible predictive capability for HET plasmas is needed.

That capability is not easy. The HET discharge couples an applied axial electric field with a predominantly radial magnetic field; electrons are magnetized and trapped in an azimuthal $E\times B$ Hall drift, ions are essentially unmagnetized and accelerate axially, and the neutral gas is
depleted by ionization within a few centimeters \citep{boeuf2017}. The resulting system of strongly coupled, stiff, nonlinear conservation laws is commonly treated with mesh-based fluid, hybrid, and kinetic
solvers \citep{boeuf2017,roy2002pop,roy2003jpp}, which require careful mesh design, time-step control, and substantial computational resources, and which assimilate experimental data only indirectly through manual calibration.

\begin{figure}[!htbp]
\centering
\begin{tikzpicture}[scale=1.0,  yscale=2.0, >=Latex]
  \fill[gray!25] (0,0.6) rectangle (6,1.4);

  \draw[thick] (0,0.6) -- (6,0.6);
  \draw[thick] (0,1.4) -- (6,1.4);

  \fill[black] (0,0.6) rectangle (0.25,1.4);
\node[above] at (0.12,1.45) {\small Anode};
\node[below] at (0.12,0.45) {\small $\hat{x}=0$};

  \draw[dashed] (6,0.5) -- (6,1.5);
  \node[below] at (6,0.45) {\small $\hat{x}=1$};
  \node[right] at (6.05,1.5) {\small exit plane};

  \node[left,align=center] at (-0.65,1.0) {};
  \node[left] at (-0.4,0.6) {\small $r_{\mathrm{in}}$};
  \node[left] at (-0.4,1.4) {\small $r_{\mathrm{out}}$};

  \draw[->,red,very thick] (1.2,1.0) -- (2.6,1.0);
  \node[red,above] at (1.9,1.05) {\small $E_x$};

  \draw[<-,blue,very thick] (3.5,1.3) -- (3.5,0.7);
  \node[blue,right] at (3.55,1.0) {\small $B_r$};

  \node[teal] at (4.6,1.0) {$\otimes$};
  \node[teal,above] at (4.6,1.1) {\small $V_{e\theta}$};

  \draw[->,thick] (5.2,1.0) -- (6.6,1.0);
  \node[above] at (5.7,1.03) {\small $\mathrm{Xe}^+$};

  \draw[->] (0,-0.1) -- (6.5,-0.1) node[right] {\small $x$ (axial)};
  \draw[->] (-1.2,0.6) -- (-1.2,1.6) node[above] {\small $r$ (radial)};
\end{tikzpicture}
\caption{Cross-section of the P5 annular discharge channel (upper half shown). Neutral xenon is injected at the anode ($\hat{x}=0$); the axial electric field $E_x$ accelerates ions toward the exit plane ($\hat{x}=1$), while the radial magnetic field $B_r$ confines the electrons, which drift azimuthally ($V_{e\theta}$, into the page) in the $\bm{E}\times\bm{B}$ direction.}
\label{fig:schematic}
\end{figure}
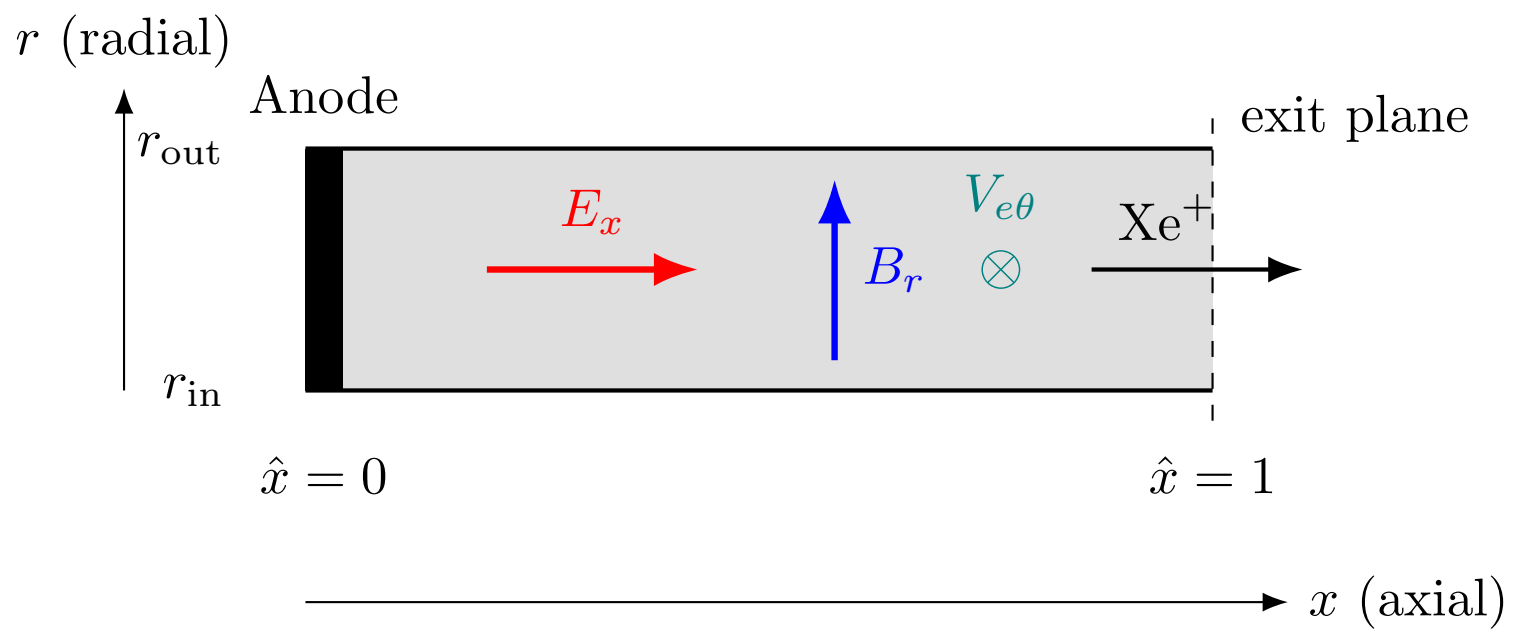

Physics-informed neural networks (PINNs) \citep{raissi2019} offer a different route. A PINN represents the solution fields with a neural network and minimizes the residuals of the governing partial differential equations, evaluated by automatic differentiation at arbitrary collocation points, together with boundary conditions and any available data \citep{karniadakis2021,cuomo2022}. The method is mesh-free, returns a continuously differentiable surrogate, and, decisively for the present work, treats sparse experimental measurements as first-class constraints alongside the physics. Within our group, PINNs have been validated on thermal cavity flows \citep{fowler2024} and applied to a simplified one-dimensional HET models \citep{burrows2024, burrows2026}; in the broader HET community, graph neural networks have also been used to accelerate the Poisson solve inside particle-in-cell simulations \citep{vigot2025}. These efforts, however, either reduce the geometry to one dimension, simplify the physics, or assist a conventional solver rather than replacing it.

\subsection{Supervised, unsupervised, and semi-supervised formulations}
\label{subsec:regimes}

Neural networks that approximate the solution of a boundary value problem are trained under two limiting regimes, distinguished by what the loss functional minimizes. In the supervised regime the loss is the distance to a known reference solution,
\begin{equation}
\mathcal{L}_{\mathrm{sup}} = \frac{1}{N}\sum_{j=1}^{N} \bigl\lVert u_{\mathrm{NN}}(\mathbf{x}_{j}) - u^{\ast}(\mathbf{x}_{j}) \bigr\rVert^{2},
\label{eq:loss_sup}
\end{equation}
where $u_{\mathrm{NN}}$ is the network approximation of the unknown fields and the labels $u^{\ast}(\mathbf{x}_{j})$ are computed beforehand by a conventional solver on a dense set of nodes covering the domain, the sum running over the $N$ nodes at which that reference solution is available. The governing equations enter only implicitly, through the labels. In the unsupervised regime the loss is the residual of the governing equations themselves,
\begin{equation}
\mathcal{L}_{\mathrm{unsup}} = \frac{1}{M}\sum_{m=1}^{M} \bigl\lVert \mathcal{N}[u_{\mathrm{NN}}](\mathbf{x}_{m}) \bigr\rVert^{2},
\label{eq:loss_unsup}
\end{equation}
where $\mathcal{N}$ denotes the governing operator evaluated at collocation points through automatic differentiation or, in discretized form, the residual $A u_{\mathrm{NN}} - b$ of the assembled linear system, and $M$ is the number of points at which the residual is evaluated, unconstrained by the availability of data since the residual follows from the operator alone. Boundary conditions are imposed as penalty terms or included in the discrete operator, and no precomputed labels or reference solution appear in the loss. The terminology is not uniform across the literature. Raissi et al. \citep{raissi2019} describe physics-informed networks as solving ``supervised learning tasks'', although in the forward problems they consider the labelled data reduce to initial and boundary values and the interior is constrained by the equation residual alone. Throughout this paper, supervised denotes training against a reference solution of the unknown fields, and unsupervised denotes training on equation residuals without such a reference, following the usage of Vigot \citep{vigot2025}.

Both regimes have been applied to the Poisson problem in Hall thruster simulation. Vigot et al. \citep{vigot2025jep,vigot2026unsup} trained graph neural networks to replace the linear solve of the discretized Poisson equation inside the AVIP particle-in-cell code, under a supervised formulation trained on potentials produced by the iterative solver and an unsupervised formulation trained on the residual of the linear system. Compared directly on the same three-dimensional configuration \citep{vigot2025}, and measured against the reference solution on the validation set, the supervised network reached a root-mean-square error of approximately 0.54, while the unsupervised network reached approximately 37.8. Burrows et al. \citep{burrows2024,burrows2026} solved a purely unsupervised forward problem, a one-dimensional quasi-neutral PINN of the discharge channel in which the loss contains the residuals of the normalized governing equations and the boundary conditions only, with no experimental measurements.

The model presented in this work occupies an intermediate position on this axis. Its composite loss, defined in Sec.~\ref{sec:loss}, carries ten governing-equation residuals and the boundary conditions together with a data term that anchors three of the unknown fields, $n_i$, $T_e$, and $V_p$, to 145 internal probe locations per field reconstructed from the data of Haas and Gallimore \citep{haas2000}. These anchors are sparse experimental points rather than dense labels supplied by a reference solver, and the remaining fields are constrained by the residuals and boundary conditions alone. In the convention adopted here the formulation is therefore semi-supervised. Composite losses of this kind, combining equation residuals with sparse or partial observations of a subset of the fields, form an established category in the physics-informed literature \citep{karniadakis2021}. In magnetized plasmas, the same construction has been used to reconstruct turbulent fields under the drift-reduced Braginskii equations from partial observations \citep{mathews2021}. The convergence metric appropriate to this training regime is discussed in Sec.~\ref{sec:results}.

The present work retains the full two-dimensional geometry and uses the PINN as a stand-alone solver that assimilates the internal probe data directly, rather than as an accelerator for a conventional code. We construct a two-dimensional, axisymmetric, steady-state PINN that solves eleven coupled plasma fields for xenon in the discharge channel of the University of Michigan and AFRL P5 Hall thruster \citep{haas2000,haas1999} at its \SI{300}{\volt}, \SI{5.4}{\ampere} operating point. The central physical contribution is the treatment of the electric potential: instead of imposing quasi-neutrality and reconstructing $\phi$ from an algebraic closure, we solve Poisson's equation explicitly in charge-separation form, $n_e = n_i(1+\eta)$, with the dimensionless charge-separation field $\eta$ produced by the network and bounded by construction. The potential is therefore self-consistent with the resolved charge densities.

The term self-consistent requires qualification, because it is often used for models in which every important variable, including the electric field, the charge densities, the velocities, and the electron energy, is computed simultaneously from the governing equations. The present model satisfies that definition only in part, and we refer to it as having \emph{limited self-consistency}. What is self-consistent is the electrostatics: $\phi$ and $\eta$ are coupled through Poisson's equation and solved together with the densities, velocities, and electron temperature in a single optimization. Several quantities are, however, imposed or only weakly constrained. The magnetic field is a prescribed analytical profile (Eq.~\eqref{eq:Bfield}) with no self-field. The discharge current is a trainable scalar held near \SI{5.4}{\ampere} by penalties, so it is imposed rather than predicted. The electron axial velocity is fixed algebraically by the current closure (Eq.~\eqref{eq:closure}), and the three electron momentum residuals carry the smallest weight assigned to any governing equation ($10^{-2}$, Table~\ref{tab:weights}). The unanchored velocity channels are bounded by their parametrization envelopes, the ion--neutral collision frequency is a prescribed constant, and the exit-plane potential is taken from the reconstructed measurements. The per-equation residuals at convergence are reported unweighted in Table~\ref{tab:residuals}, so that the degree to which each equation is satisfied can be read directly. Section~\ref{sec:consistency} relaxes several of these restrictions: it replaces the analytical field by a measured one, determines the electron axial velocity from the momentum equations, and enforces the global balances and the discharge current as integrated constraints.

The formulation is made trainable by rescaling the Poisson residual by its stiffness coefficient $\alpha_p = e\,n^{*}L^{2}/(\varepsilon_0\phi_{\mathrm{ref}}) \approx 7.24\times10^{7}$. The network is trained against, and compared with, the internal probe measurements of Haas and Gallimore \citep{haas2000} on ion density, electron temperature, and plasma potential at five radial positions; since these points enter the objective, the comparison measures in-sample agreement. Two further studies probe what the solution rests on: a comparison between the reference electron transport closure and a classical-transport configuration, which shows how weakly the measured fields constrain the electron momentum balance; and a set of held-out tests, in which one or two radial measurement lines are withdrawn from the objective, which separates the fields supplied by the governing equations from those supplied by the data. The held-out tests yield the central finding of this work: the plasma potential is insensitive to which measurement lines are retained, whereas the ion-density reconstruction is highly sensitive to the placement of the measurements. The tests do not isolate how much of the robustness of the potential is due to Poisson's equation rather than to its boundary values and the remaining anchors. That sensitivity defines where the method can be trusted and which probe locations matter most, and it is reported here as a result in its own right rather than as a caveat on a predictive solver.

A good fit to the measurements does not establish that the governing equations are satisfied in an integral sense, and the second part of the paper tests the reference solution against requirements that the data term does not see: the global particle and current balances, the electron current implied by the electron momentum equation, the force balance along the magnetic field, and the axial ion momentum budget. Each diagnostic exposes a limitation of the reference configuration, and the limitations are addressed by successive groups of corrections, several of which change more than one component at once. The sequence leads to a solution that conserves particles, holds the discharge current near its measured value and matches the potential and temperature measurements within their uncertainties, and it identifies the assumptions and measurements that still prevent full self-consistency.

The remainder of the paper is organized as follows. Section~2 presents the physical model: governing equations, charge-separation form of Poisson's equation, magnetic field and transport model, normalization, and boundary conditions. Section~3 details the network architecture, the hard encoding of boundary conditions, the composite loss, and the training procedure. Section~4 describes the experimental dataset and its assimilation. Section~5 reports the reference solution, its validation against experiment and comparison with a published hybrid particle-in-cell simulation of the same thruster, the convergence behavior and training cost, the effect of the anomalous transport term, the complete set of converged fields, and the held-out tests. Section~6 presents the physical-consistency analysis and the resulting model refinements. Section~7 concludes with limitations and future work.

\section{Physical Model}

\subsection{Geometry and modeling assumptions}
The computational domain is the annular discharge channel of the P5, a 5-kW-class laboratory Hall thruster developed jointly by the University of Michigan and the Air Force Research Laboratory \citep{haas2000}. The channel has length $L = \SI{38}{\milli\meter}$ and spans inner and outer radii $r_{\mathrm{in}} = \SI{60}{\milli\meter}$ and $r_{\mathrm{out}} = \SI{85}{\milli\meter}$, a \SI{25}{\milli\meter} width \citep{haas1999,haas2000}. Axial and radial coordinates are both normalized by $L$, so that $\hat{x} = x/L \in [0,1]$ from anode ($\hat{x}=0$) to exit plane ($\hat{x}=1$) and $\hat{r} = r/L \in [\hat{r}_{\mathrm{in}},\hat{r}_{\mathrm{out}}] = [1.579,\,2.237]$ across the channel width. The problem is treated as steady-state and axisymmetric ($\partial_\theta=0$), with the azimuthal electron drift $V_{e\theta}$ retained as a field variable.

Three xenon species are modeled as interpenetrating fluids, following the assumptions standard in the multifluid HET literature \citep{roy2002pop,roy2003jpp}: (i) singly charged ions, cold and unmagnetized, the ion Larmor radius greatly exceeding the channel length so that the magnetic force on ions is neglected and the cold-ion approximation removes the ion pressure term; (ii) electrons, magnetized and warm, described by continuity, the three components of momentum including the Hall coupling, and an energy equation for $T_e$; (iii) neutrals, advected at a constant axial drift velocity $V_n = \SI{450}{\meter\per\second}$ ($0.151$ normalized) with no radial drift. The model comprises eleven physical fields: the potential $\phi$; the densities $n_i$, $n_e$, $n_n$; the ion velocities $V_{ix}$, $V_{ir}$; the electron velocities $V_{ex}$, $V_{er}$, $V_{e\theta}$; the electron temperature $T_e$; and the relative charge separation $\eta$. Two of these are not independent: $n_e$ follows from $n_i$ and $\eta$ through Eq.~\eqref{eq:chargesep}, and $V_{ex}$ is fixed by the current closure of Eq.~\eqref{eq:closure}.

The applied magnetic field is taken as purely radial, $\mathbf{B} = B_r(\hat{x})\,\hat{\mathbf{r}}$, with a Gaussian axial profile
\begin{equation}
B_r(\hat{x}) = B_{\max}\,\exp\!\left[-\frac{(\hat{x}-x_B)^2}{\sigma_B^{2}}\right],
\label{eq:Bfield}
\end{equation}
consistent with the bell-shaped radial-field profiles measured in the P5 discharge channel \citep{peterson2002}. The reference run uses $B_{\max} = \SI{0.02}{\tesla}$, with peak location $x_B = 0.75$ and width parameter $\sigma_B = 0.30$. Because $\mathbf{B}$ is purely radial, the Lorentz force has no radial component: $(\mathbf{V}_e\times\mathbf{B})_r = 0$ identically, and the field couples only the axial and azimuthal electron motions. The electric field is not similarly restricted: Poisson's equation is solved in two dimensions, so that $E_r= -\partial\phi/\partial r$ is retained. The radial component $B_r$ is moreover taken uniform across the channel width, a one-dimensional idealization of the measured P5 topography \citep{peterson2002} that carries no radial information; all radial structure in the solution is therefore driven by the radial electric field, the data, and the boundary conditions. Section~\ref{subsec:bfield} replaces this profile by a two-dimensional field fitted to measurements.

\subsection{Governing Equations}

All equations below are written in the dimensionless variables of Sec.~\ref{sec:norm} and at steady state. The PINN minimizes the residual of each equation, obtained by moving all terms to one side and evaluating the result by automatic differentiation at the collocation points (Sec.~\ref{sec:loss}). The system comprises Poisson's equation in charge-separation form, three continuity equations, two ion momentum equations, three electron momentum equations, the electron energy equation, and a current-closure constraint.

\subsubsection*{Poisson's equation in charge-separation form}
The electron density is written as
\begin{equation}
n_e = n_i\,(1+\eta), \qquad |\eta| \le \eta_{\max},
\label{eq:chargesep}
\end{equation}
where $\eta(\hat{x},\hat{r})$ is the relative charge separation, a network output bounded by construction with $\eta_{\max} = 10^{-4}$. Combining Gauss's law with $\mathbf{E}=-\nabla\phi$ gives Poisson's equation, $\nabla^2\phi = -(e/\varepsilon_0)(n_i-n_e)$, which in cylindrical coordinates and normalized form becomes
\begin{equation}
\frac{1}{\alpha_p}\left(\frac{\partial^2\phi}{\partial x^2} + \frac{\partial^2\phi}{\partial r^2} + \frac{1}{r}\frac{\partial\phi}{\partial r}\right) = n_i\,\eta,
\label{eq:poisson}
\end{equation}
with the Poisson stiffness coefficient
\begin{equation}
\alpha_p = \frac{e\,n^{*}L^{2}}{\varepsilon_0\,\phi_{\mathrm{ref}}} = 7.242\times10^{7}.
\label{eq:alphap}
\end{equation}
Two features make Eq.~\eqref{eq:poisson} trainable. First, the unscaled form $\nabla^2\phi + \alpha_p(n_i-n_e)$, evaluated with two independently parametrized densities, contains terms of order $\alpha_p\times10^{-2}\approx10^{6}$ at initialization; squared in the loss they exceed every other term by many orders of magnitude and saturate the optimizer. Writing $n_i-n_e=-n_i\eta$ replaces the difference of two independent outputs by a single order-one network output $\eta/\eta_{\max}$, and dividing through by $\alpha_p$ keeps both sides of order $n_i\eta_{\max}$. Second, the corresponding residual is rescaled by $1/\eta_{\max}$ (Sec.~\ref{sec:loss}) so that deviations of $\eta$ from the value demanded by the Laplacian are penalized on the natural scale of the charge-separation bound. The bound is not arbitrary: with $n_i\sim5\times10^{-2}$, $\alpha_p n_i\eta_{\max}\approx4\times10^{2}$, the magnitude of the Laplacian it must balance. In the limit $\eta\to0$ the formulation reduces to $n_e = n_i$;quasi-neutrality is therefore imposed as a bound rather than as an exact value, the magnitude, sign, and spatial distribution of $\eta$ within that bound are determined by training, and the potential is obtained from the resolved charge densities rather than from an algebraic closure.

\subsubsection*{Continuity}
The following mass conservation equations for ions, neutrals, and electrons, respectively, are considered
\begin{align}
\frac{\partial(n_i V_{ix})}{\partial x} + \frac{\partial(n_i V_{ir})}{\partial r} + \frac{n_i V_{ir}}{r} &= S_i,
\label{eq:ioncont}\\
V_n\,\frac{\partial n_n}{\partial x} &= -\,S_i,
\label{eq:neutcont}\\
\frac{\partial(n_e V_{ex})}{\partial x} + \frac{\partial(n_e V_{er})}{\partial r} + \frac{n_e V_{er}}{r} &= S_i,
\label{eq:eleccont}
\end{align}
with the ionization source $S_i$ defined in Eq.~\eqref{eq:source}. The neutral sink and the ion and electron sources share the same term, so that the particles removed from the neutral population are exactly those added to the charged populations. The neutral equation carries no radial transport, the neutrals being given a prescribed constant axial velocity.

\subsubsection*{Ion momentum}
With cold, unmagnetized ions subjected to the electric force, electron collisional momentum exchange at the classical rate $\nu_c = \nu_{ei}+\nu_{en}$, ion--neutral friction, and the momentum injected by ionization at the neutral drift velocity, the axial and radial balances are
\begin{align}
\frac{\partial(n_i V_{ix}^{2})}{\partial x} + \frac{\partial(n_i V_{ix}V_{ir})}{\partial r} + \frac{n_i V_{ix}V_{ir}}{r} + n_i\,\frac{\partial\phi}{\partial x}
&= n_i\frac{m_e}{m_i}\,\hat{\nu}_c\,(V_{ex}-V_{ix}) \nonumber\\
&\quad - n_i\frac{m_n}{m_i}\,\hat{\nu}_{in}\,(V_{ix}-V_n) + S_i V_n,
\label{eq:ionmomx}\\[4pt]
\frac{\partial(n_i V_{ix}V_{ir})}{\partial x} + \frac{\partial(n_i V_{ir}^{2})}{\partial r} + \frac{n_i V_{ir}^{2}}{r} + n_i\,\frac{\partial\phi}{\partial r}
&= n_i\frac{m_e}{m_i}\,\hat{\nu}_c\,(V_{er}-V_{ir}) \nonumber\\
&\quad - n_i\frac{m_n}{m_i}\,\hat{\nu}_{in}\,V_{ir}.
\label{eq:ionmomr}
\end{align}
The ion drag uses the classical rate $\nu_c$ and excludes the anomalous contribution introduced below, which represents cross-field electron transport rather than a physical friction between ions and electrons.

\subsubsection*{Electron momentum (Hall coupling)}
With $\omega_{ce} = eB_r/m_e$ the electron cyclotron frequency associated with the radial field of Eq.~\eqref{eq:Bfield}, the three components read
\begin{align}
\frac{m_e}{m_i}\left[n_e V_{ex}\frac{\partial V_{ex}}{\partial x} + n_e V_{er}\frac{\partial V_{ex}}{\partial r}\right]
&= n_e\,\frac{\partial\phi}{\partial x} + \frac{m_e}{m_i}\,\hat{\omega}_{ce}\,n_e V_{e\theta} - \frac{m_e}{m_i}\,\hat{\nu}\, n_e V_{ex},
\label{eq:emomx}\\
\frac{m_e}{m_i}\left[n_e V_{ex}\frac{\partial V_{er}}{\partial x} + n_e V_{er}\frac{\partial V_{er}}{\partial r}\right]
&= n_e\,\frac{\partial\phi}{\partial r} - \frac{m_e}{m_i}\,\hat{\nu}\, n_e V_{er},
\label{eq:emomr}\\
\frac{m_e}{m_i}\left[n_e V_{ex}\frac{\partial V_{e\theta}}{\partial x} + n_e V_{er}\frac{\partial V_{e\theta}}{\partial r}\right]
&= -\,\frac{m_e}{m_i}\,\hat{\omega}_{ce}\,n_e V_{ex} - \frac{m_e}{m_i}\,\hat{\nu}\, n_e V_{e\theta},
\label{eq:emomtheta}
\end{align}
where $\hat{\omega}_{ce} = \omega_{ce}\tau^{*}$ and $\hat{\nu} = \nu\tau^{*}$. The mass ratio appears in every term except the electric force because the velocity scale of Table~\ref{tab:norm} is based on the ion mass. Because $\mathbf{B}=B_r\hat{\mathbf{r}}$, $B_r$ couples only the axial and azimuthal electron motions, and the radial electric force is retained in Eq.~\eqref{eq:emomr}. Equation~\eqref{eq:emomtheta} is the Hall-drift equation; the azimuthal velocity it produces feeds back into the axial balance, Eq.~\eqref{eq:emomx}, which is the mechanism by which the magnetic barrier suppresses axial electron mobility and allows the plasma to maintain the applied potential drop. The electron pressure gradient does not appear in the three components of the reference model; with $n_e$ and $T_e$ both resolved, an isotropic closure $p_e=n_eT_e$ is available, and it is added in Sec.~\ref{subsec:stageA}.

\subsubsection*{Electron energy}
With $\varepsilon_i = E_I/\phi_{\mathrm{ref}}$ the normalized xenon first ionization energy, equal to one because $\phi_{\mathrm{ref}} = E_I$, the steady-state electron energy balance is
\begin{align}
\frac{5}{2}\left[\frac{\partial(n_e V_{ex}T_e)}{\partial x}
+ \frac{\partial(n_e V_{er}T_e)}{\partial r}
+ \frac{n_e V_{er}T_e}{r}\right]
&= n_e V_{ex}\frac{\partial\phi}{\partial x} + n_e V_{er}\frac{\partial\phi}{\partial r}
+ 3\frac{m_e}{m_i}n_e\hat{\nu}_{ei}(T_i-T_e) \nonumber\\
&\quad + 3\frac{m_e}{m_n}n_e\hat{\nu}_{en}(T_n-T_e)
+ S_i\!\left(\tfrac{3}{2}T_e + \alpha_E \varepsilon_i\right),
\label{eq:energy}
\end{align}
comprising enthalpy convection on the left and, on the right, Ohmic heating by the electric field, collisional temperature exchange with ions and neutrals, and the ionization energy term of Roy and Pandey \citep{roy2002pop} and Burrows et al.\ \citep{burrows2024}. Energy transport is convective only: no conductive heat flux appears. The heavy-species temperatures are $T_i = T_n = \SI{0.025}{\electronvolt}$ (room temperature), and the ionization cost factor is $\alpha_E = 2.5$.

\subsubsection*{Current closure}
Steady state requires $\nabla\cdot\mathbf{j}=0$, which constrains the current integrated over the channel cross-section rather than the local current density. The present formulation goes further and treats the axial current density $J_T = n_i V_{ix} - n_e V_{ex}$ as a single scalar, uniform in both coordinates; this is an additional modeling assumption rather than a consequence of steady state. The electron axial velocity is fixed by the current balance,
\begin{equation}
V_{ex} = \frac{n_i V_{ix} - J_T}{\max(n_e,\epsilon)},
\label{eq:closure}
\end{equation}
where $\epsilon = 10^{-6}$ guards the division, so that $n_i V_{ix} - n_e V_{ex} = J_T$ holds by construction wherever the density guard is inactive. The dimensionless current relates to amperes through $J_T[\mathrm{A}] = J_T\,e\,n^{*}V^{*}A_{\mathrm{ch}}$, with channel cross-section $A_{\mathrm{ch}} = \pi(r_{\mathrm{out}}^2 - r_{\mathrm{in}}^2) = 1.139\times10^{-2}\ \mathrm{m^2}$, giving \SI{182.4}{\ampere} per unit $J_T$; the target $J_T = 0.0296$ thus corresponds to \SI{5.4}{\ampere}. The scalar $J_T$ is itself trainable and is held near the operating point by penalties (Sec.~\ref{sec:param}).

\subsubsection*{Ionization, collisions, and anomalous transport}
The ionization source is
\begin{equation}
S_i = k_i(T_e)\,n_n\,n_e,
\label{eq:source}
\end{equation}
with the single-ionization rate coefficient of Roy and Pandey \citep{roy2002jpp}, a third-order polynomial fitted to experimental xenon data,
\begin{equation}
k_i^{\mathrm{phys}}(T_e) =
\left(1.9435\times10^{-5}\,T_e^{3} - 6.8\times10^{-3}\,T_e^{2}
+ 0.6705\,T_e - 1.6329\right)\times10^{-14}
\quad[\mathrm{m^{3}\,s^{-1}},\ T_e\ \mathrm{in\ eV}].
\label{eq:ki}
\end{equation}
Below \SI{5}{\electronvolt} the raw polynomial becomes negative and unphysical, so $k_i$ is smoothly tapered by a sigmoid factor $\sigma\!\left[(T_e - T_{e,0})/\Delta T\right]$ centered at $T_{e,0} = \SI{5}{\electronvolt}$ with width $\Delta T = \SI{0.35}{\electronvolt}$; the temperature entering the polynomial is itself floored through a softplus, $T_e^{\mathrm{eff}} = T_{e,\mathrm{floor}} + \mathrm{softplus}[\beta(T_e - T_{e,\mathrm{floor}})]/\beta$, with $T_{e,\mathrm{floor}} = \SI{3.5}{\electronvolt}$ and $\beta = 8.0$. Collision frequencies follow standard expressions:
\begin{gather}
\nu_{ei} = 2.91\times10^{-6}\,n_e^{\mathrm{cm}}\,\ln\Lambda\;T_e^{-3/2},\nonumber
\qquad
\nu_{en} = n_n\,\sigma_{en}\sqrt{\frac{8\,T_e e}{\pi m_e}},
\\
\nu_B = \alpha_B\,\omega_{ce},
\qquad
\nu_{in} = 1.8\times10^{6}\ \mathrm{s^{-1}}\ \text{(const.)},
\label{eq:collisions}
\end{gather}
with $n_e^{\mathrm{cm}}$ the electron density in $\mathrm{cm^{-3}}$, the Coulomb logarithm $\ln\Lambda = 10$, and the electron--neutral cross section $\sigma_{en} = 2.5\times10^{-19}\,\mathrm{m^{2}}$. The momentum-transfer frequency entering the electron momentum equations is $\nu = \nu_{ei}+\nu_{en}+\nu_B$. The third term is an anomalous contribution of Bohm form with $\alpha_B = 1/200$: classical electron--ion and electron--neutral collisions alone give a cross-field mobility far below that measured in these devices, and the Bohm term is the standard means of representing the difference within a fluid formulation \citep{roy2002pop}. Its effect on the Hall parameter $\Omega = \omega_{ce}/\nu$ is examined in Sec.~\ref{sec:bohm}. The ion--neutral frequency is the only collision rate in the model that is prescribed rather than evaluated from the local fields.

\subsection{Normalization}
\label{sec:norm}
The nondimensionalization follows the scheme of Roy and Pandey \citep{roy2002pop,roy2002jpp}. The temperature and potential references are set by the xenon first ionization energy, $T^{*} = E_I = \SI{12.1}{\electronvolt}$ (so $\phi_{\mathrm{ref}} = \SI{12.1}{\volt}$); the velocity scale is set by the ionization energy, $V^{*} = \sqrt{E_I e/m_i} = \SI{2981.9}{\meter\per\second}$; the flux scale is taken as $\Gamma^{*} = \SI{e23}{\per\meter\squared\per\second}$, a reference value of the scheme rather than the flux of the operating point, which is about forty times smaller; densities follow as $n^{*} = \Gamma^{*}/V^{*} = \SI{3.354e19}{\per\meter\cubed}$; and the collision-frequency scale is $\nu^{*} = \sigma^{*}\Gamma^{*}$ with $\sigma^{*} = \sigma_0\,(m_i/m_e)^{1/2}$ and $\sigma_0 = \SI{3.6e-20}{\meter\squared}$ for xenon.

Three elements are extensions introduced in the present work and must be distinguished from the original scheme. First, whereas Roy and Pandey adopt the collisional length $l_0 = V^{*}/\nu^{*}$ as fundamental scale, we normalize lengths by the channel length $L = \SI{0.038}{\meter}$, the natural scale of the computational domain; the collision and cyclotron frequencies entering Eqs.~\eqref{eq:ionmomx}--\eqref{eq:energy} are accordingly nondimensionalized by the transit time, $\hat{\nu}=\nu\tau^{*}$ with $\tau^{*}=L/V^{*}$. Second, solving Poisson's equation explicitly introduces the stiffness coefficient $\alpha_p$ of Eq.~\eqref{eq:alphap}, which has no counterpart in a quasi-neutral formulation. Third, the source and rate scales, $S^{*} = n^{*}V^{*}/L$ and $k^{*} = V^{*}/(L\,n^{*})$, are defined to nondimensionalize Eqs.~\eqref{eq:source} and \eqref{eq:ki} consistently with the channel-length normalization. Table~\ref{tab:norm} collects the normalization scales.

\begin{table}[H]
\centering
\caption{Normalization scales. The entries above the rule follow Roy and Pandey \citep{roy2002pop,roy2002jpp}; $\alpha_p$, $S^{*}$, $k^{*}$, and the use of $L$ as length scale are extensions of the present work.}
\label{tab:norm}
\setlength{\tabcolsep}{13pt}
\begin{tabular}{llll}
\toprule
Quantity & Symbol & Value & Unit \\
\midrule
Ionization energy & $T^{*}=E_I$ & $12.1$ & \si{\electronvolt} \\
Potential scale & $\phi_{\mathrm{ref}}$ & $12.1$ & \si{\volt} \\
Velocity scale & $V^{*}$ & $2981.9$ & \si{\meter\per\second} \\
Flux scale & $\Gamma^{*}$ & $1.000\times10^{23}$ & \si{\per\meter\squared\per\second} \\
Density scale & $n^{*}=\Gamma^{*}/V^{*}$ & $3.354\times10^{19}$ & \si{\per\meter\cubed} \\
Ref.\ cross section & $\sigma_0$ & $3.600\times10^{-20}$ & \si{\meter\squared} \\
Collision-freq.\ scale & $\nu^{*}=\sigma^{*}\Gamma^{*}$ & $1.761\times10^{6}$ & \si{\per\second} \\
\midrule
Channel length & $L$ & $0.038$ & \si{\meter} \\
Time scale & $\tau^{*}=L/V^{*}$ & $1.274\times10^{-5}$ & \si{\second} \\
Poisson scale & $\alpha_p=\dfrac{e\,n^{*}L^{2}}{\varepsilon_0\phi_{\mathrm{ref}}}$ & $7.242\times10^{7}$ & \\
Source scale & $S^{*}=n^{*}V^{*}/L$ & $2.632\times10^{24}$ & \si{\per\meter\cubed\per\second} \\
Rate scale & $k^{*}=V^{*}/(L\,n^{*})$ & $2.340\times10^{-15}$ & \si{\meter\cubed\per\second} \\
\bottomrule
\end{tabular}
\end{table}

\subsection{Boundary conditions}
\label{sec:bc}
Table~\ref{tab:bc} lists the boundary conditions of the reference configuration with their dimensional and normalized values and their enforcement type. \emph{Hard} conditions are satisfied exactly by the network parametrization for every value of the weights; \emph{soft} conditions enter the loss as penalties. The anode potential is the \SI{300}{\volt} applied discharge voltage. The exit-plane potential is set from the reconstructed Haas--Hofer--Gallimore P5 HARP data \citep{haas1999}, using the mean of the near-exit reconstructed plasma potential over $x/L \ge 0.95$, giving \SI{116.2}{\volt}. Because the internal probe measurements begin approximately \SI{10}{\milli\meter} downstream of the anode face \citep{haas2000}, the anode values of $T_e$ and $n_i$ are prescribed rather than taken from the internal maps directly: the near-anode electron temperature $T_e = \SI{2.0}{\electronvolt}$ follows the value reported close to the anode by Roy and Pandey \citep{roy2002pop}, and the anode ion density $n_i = \SI{4.64e17}{\per\meter\cubed}$ is chosen to match the order of magnitude of the first available Haas--Gallimore near-anode density measurements at $x = \SI{10}{\milli\meter}$. The neutral density is set to unity (its inlet reference) at the anode, where propellant is injected; with the prescribed neutral velocity this corresponds to an inflow 6.1 times the measured anode flow, which Sec.~\ref{subsec:global} corrects, and the ion axial velocity is fixed to a small positive floor at the anode, $V_{ix,\min}=0.025$ in normalized units, consistent with the near-zero axial velocity reported close to the anode by Roy and Pandey \citep{roy2002pop}. Two velocity fields, $V_{e\theta}$ and $V_{er}$, are zeroed at their boundaries both by their parametrization envelopes and by soft penalties; the envelopes additionally force $V_{e\theta}$ to vanish at the exit and $V_{ir}$ at the anode, which are structural consequences of the parametrization rather than physical conditions and are listed so that the full set of constraints is visible.

\begin{table}[H]
\centering
\caption{Boundary conditions of the reference run (normalized values in parentheses). Rows marked $^{\dagger}$ are consequences of the output parametrizations of Sec.~\ref{sec:param} rather than conditions imposed for physical reasons.}
\label{tab:bc}
\begin{tabular}{llll}
\toprule
Field & Location & Value & Enforcement \\
\midrule
$\phi$ & anode ($\hat{x}{=}0$) & \SI{300}{\volt} (24.79) & hard \\
$\phi$ & exit ($\hat{x}{=}1$) & \SI{116.2}{\volt} (9.603) & hard \\
$T_e$ & anode & \SI{2.0}{\electronvolt} (0.1653) & hard \\
$n_i$ & anode & $4.64\times10^{17}\ \mathrm{m^{-3}}$ (0.01384) & hard \\
$n_n$ & anode & $1.0$ & hard \\
$V_{ix}$ & anode & \SI{74.5}{\meter\per\second} (0.025) & hard floor \\
$V_{e\theta}$ & anode & 0 & envelope (hard) and penalty \\
$V_{e\theta}$ & exit$^{\dagger}$ & 0 & envelope (hard) \\
$V_{ir}$ & walls & 0 & penalty \\
$V_{ir}$ & anode$^{\dagger}$ & 0 & envelope (hard) \\
$V_{er}$ & walls & 0 & envelope (hard) and penalty \\
$J_T$ & --- & \SI{5.4}{\ampere} target & scalar penalties \\
\bottomrule
\end{tabular}
\end{table}

\section{PINN Architecture}
\label{sec:arch}
\subsection{Network architecture}

The solver is implemented in PyTorch \citep{paszke2019} in double precision (float64), chosen for margin in the Poisson residual, where differences of order $10^{-6}$ must be resolved. A single fully connected network maps the coordinates $(x,r)\in\mathbb{R}^{2}$ to ten raw output channels through four hidden layers of 128 neurons with $\tanh$ activations, a linear output head, and Xavier-uniform initialization. The $\tanh$ activation is chosen because the Poisson residual requires nonvanishing, smooth second derivatives everywhere in the domain, which $C^{\infty}$ bounded activations provide.

The ten raw channels map to the eleven physical fields $(\phi, n_i, V_{ix}, V_{ir}, n_n, n_e, V_{ex}, V_{er}, V_{e\theta}, T_e, \eta)$ through the hard-constraint transformations of Sec.~\ref{sec:param}. Two of the eleven are not free outputs: the electron density $n_e$ is reconstructed from the charge-separation identity of Eq.~\eqref{eq:chargesep}, and the electron axial velocity is fixed by the current closure of Eq.~\eqref{eq:closure}. The architecture provides a raw channel for a bounded additive correction to $V_{ex}$, but its amplitude is set to zero in the reference configuration, so nine channels carry the solution and the tenth is present but inactive.

\subsection{Hard-constraint parametrization}
\label{sec:param}

Hard boundary conditions are encoded as differentiable output transformations so that no admissible weight configuration can violate them. This removes the corresponding penalty competition from the loss and, equally important, excludes the degenerate solutions ($n_e\to0$, collapse to an $r$-independent profile) that satisfy the residuals trivially. Writing $\xi\in[0,1]$ and $\rho\in[0,1]$ for the axial and radial coordinates normalized to the unit interval over the channel and $\hat{u}_k$ for the $k$-th raw network channel, the parametrized fields are
\begin{align}
\phi &= \phi_{\mathrm{base}} + \xi(1-\xi)\,A_\phi\,\tanh(\hat{u}_0),
\label{eq:phi_param}\\
n_i &= n_{i,\mathrm{in}}\,\exp\!\big[a_{\mathrm{seed}}\,4\xi(1-\xi) + \xi\,\hat{u}_1\big],
\label{eq:ni_param}\\
V_{ir} &= 0.005\,\xi\,\tanh(\hat{u}_3),
\label{eq:vir_param}\\
V_{er} &= 0.008\,\rho(1-\rho)\,\tanh(\hat{u}_5),
\label{eq:ver_param}\\
V_{e\theta} &= 8000\,\xi(1-\xi)\,\tanh(\hat{u}_6),
\label{eq:vetheta_param}\\
\eta &= \eta_{\max}\,\tanh(\hat{u}_7/3),
\label{eq:eta_param}\\
T_e &= T_{e,\mathrm{an}} + \xi\,\mathrm{softplus}(\hat{u}_8),
\label{eq:te_param}\\
V_{ix} &= V_{ix,\mathrm{min}} + \xi\,\mathrm{softplus}(\hat{u}_9),
\label{eq:vix_param}
\end{align}
where $\phi_{\mathrm{base}} = \phi_{\mathrm{left}} + (\phi_{\mathrm{right}}-\phi_{\mathrm{left}})\,\xi$ linearly interpolates the two axial Dirichlet values, $A_\phi = 25$ scales the interior correction, and the factor $\xi(1-\xi)$ makes the correction vanish at the anode and exit. The density parametrization~\eqref{eq:ni_param} guarantees positivity and the exact anode value $n_{i,\mathrm{in}}$; the seed term $a_{\mathrm{seed}}\,4\xi(1-\xi)$, with $a_{\mathrm{seed}} = \ln\max(n_i^{\mathrm{seed}}/n_{i,\mathrm{in}},\,1)$ and $n_i^{\mathrm{seed}} = 0.060$, biases the initial profile toward a central density bump so that the constant solution is not a stationary point of the loss at initialization. The envelopes and amplitude constants of Eqs.~\eqref{eq:vir_param}--\eqref{eq:vetheta_param} bound the unanchored velocity channels, $|V_{ir}|<0.005$, $|V_{er}|\le0.002$ (about \SI{6}{\meter\per\second}), and $|V_{e\theta}|\le2000$ (about \SI{6.0e6}{\meter\per\second}), which prevents a channel with no data anchor from growing without physical justification to relieve a residual elsewhere in the system. The azimuthal bound places the $\bm{E}\times\bm{B}$ drift sustained by the discharge well inside the permitted range. The velocity form~\eqref{eq:vix_param} enforces the anode floor $V_{ix,\mathrm{min}} = 0.025$, the electron temperature~\eqref{eq:te_param} is anchored to its anode value $T_{e,\mathrm{an}} = \SI{2.0}{\electronvolt}$, and the charge-separation field~\eqref{eq:eta_param} is bounded within $[-\eta_{\max},\eta_{\max}]$ by construction. The neutral density uses an anode-referenced depletion form,
\begin{equation}
n_n = n_{n,\mathrm{in}}\,\exp(-D), \qquad
D = \mathrm{ReLU}\!\big[\mathrm{softplus}(\hat{u}_4 - \hat{u}_4|_{x=0}) - \ln 2\big],
\label{eq:nn_param}
\end{equation}
so that $D=0$ at the anode (giving $n_n = n_{n,\mathrm{in}}$) and $0 < n_n \le n_{n,\mathrm{in}}$ everywhere, enforcing monotone propellant depletion from the inlet. Finally, the discharge current is a trainable scalar, $J_T = J_{\mathrm{init}}\exp[0.3\tanh(J_{\mathrm{raw}})]$, confined to a multiplicative range of $e^{\pm0.3}$ about its initial value, and held near \SI{5.4}{\ampere} by two scalar penalties, one on the normalized value and one on the value in amperes (Table~\ref{tab:weights}).

\begin{table}[ht]
\centering
\caption{Network and training configuration of the reference run.}
\label{tab:network}
\begin{tabular}{llll}
\toprule
Input & $(x,r)\in\mathbb{R}^{2}$ & Optimizer & Adam, lr $=10^{-4}$ \\
Raw outputs & 10 (9 active) & LR schedule & ReduceLROnPlateau (0.5, p$=$300) \\
Hidden layers & 4 & Precision & float64 \\
Neurons/layer & 128 & Collocation & 10000 int.\ $+$ 1000 inlet $+$ 1000 wall \\
Data points & 145 per field & Epochs & 15,000\\
Activation & $\tanh$ & Init & Xavier uniform \\
\bottomrule
\end{tabular}
\end{table}

\subsection{Composite loss}
\label{sec:loss}
The training objective is the weighted sum of three contributions: the physics residuals, the boundary penalties, and the experimental data terms,
\begin{equation}
\mathcal{L} \;=\; \mathcal{L}_{\mathrm{PDE}} \;+\; \mathcal{L}_{\mathrm{BC}} \;+\; \mathcal{L}_{\mathrm{data}}.
\label{eq:loss_total}
\end{equation}
Each contribution is itself a weighted sum of mean-squared terms over the relevant sampling set, with a single family of weights $w$. The physics term collects the governing-equation residuals $r_k$ over the $N_c$ interior collocation points,
\begin{equation}
\mathcal{L}_{\mathrm{PDE}} \;=\; \sum_{k} w_k\,\frac{1}{N_c}\sum_{i=1}^{N_c} r_k(\hat{x}_i,\hat{r}_i)^2,
\label{eq:loss_pde}
\end{equation}
where $k$ runs over Poisson's equation, the three continuity equations, the five momentum components, and the electron energy equation. The boundary term collects the soft boundary conditions, each the mean squared violation of a prescribed value $g_b$ over its boundary set $\mathcal{B}_b$ of $N_b$ points,
\begin{equation}
\mathcal{L}_{\mathrm{BC}} \;=\; \sum_{b} w_b\,\frac{1}{N_b}\sum_{i\in\mathcal{B}_b} \big(u_b(\hat{x}_i,\hat{r}_i) - g_b\big)^2,
\label{eq:loss_bc}
\end{equation}
where $u_b$ is the constrained field and $g_b$ is the constant prescribed value. The data term anchors three fields to the reconstructed Haas--Gallimore profiles (Sec.~\ref{sec:data}) over the $N_d$ measurement points,
\begin{equation}
\mathcal{L}_{\mathrm{data}} \;=\; w_{n_i}\,\mathcal{L}^{\mathrm{data}}_{n_i} \;+\; w_{T_e}\,\mathcal{L}^{\mathrm{data}}_{T_e} \;+\; w_{V_p}\,\mathcal{L}^{\mathrm{data}}_{V_p}.
\label{eq:loss_data}
\end{equation}
Because the ion density spans more than an order of magnitude across the channel, its term uses the log-density mean squared error, while the electron temperature and plasma potential use the mean squared error normalized by their reference scales,
\begin{align}
\mathcal{L}^{\mathrm{data}}_{n_i} &= \frac{1}{N_d}\sum_{j=1}^{N_d}
\big(\ln n_{i,j}^{\mathrm{pred}} - \ln n_{i,j}^{\mathrm{data}}\big)^2,
\label{eq:loss_ni}\\[2pt]
\mathcal{L}^{\mathrm{data}}_{T_e} &= \frac{1}{N_d}\sum_{j=1}^{N_d}
\left(\frac{T_{e,j}^{\mathrm{pred}} - T_{e,j}^{\mathrm{data}}}{T^{*}}\right)^2,
\label{eq:loss_te}\\[2pt]
\mathcal{L}^{\mathrm{data}}_{V_p} &= \frac{1}{N_d}\sum_{j=1}^{N_d}
\left(\frac{\phi_j^{\mathrm{pred}} - \phi_j^{\mathrm{data}}}{\phi_{\mathrm{ref}}}\right)^2,
\label{eq:loss_vp}
\end{align}
where the subscript $j$ runs over the measurement points and $T^{*}$, $\phi_{\mathrm{ref}}$ are the temperature and potential scales of Table~\ref{tab:norm}. Four weak smoothness regularizers, on $n_i$, $n_n$, $\phi$, and $T_e$, damp high-frequency artifacts; they carry the smallest weights in the hierarchy and are not part of the physical model.

Two rescalings precede the weighting. The first conditions each governing residual so that the terms are of comparable magnitude at initialization: the ion continuity residual is scaled by $0.6101$, the electron energy residual by $0.9652$, and the ion momentum residuals by $0.7937$. The second is specific to Poisson's equation, whose residual is rescaled by $1/\eta_{\max}$ for the reason given in Sec.~2.2; the reported Poisson residual carries this factor and is therefore not directly comparable with the raw residuals of the other equations. Table~\ref{tab:weights} gives the full weight set; the data weights $(w_{n_i},w_{T_e},w_{V_p}) = (10,5,30)$ apply to all five instrumented radii.

\begin{table}[ht]
\centering
\caption{Loss weights of the reference run.}
\label{tab:weights}
\begin{tabular}{lr lr lr}
\toprule
Poisson & 100 & ion cont. & 200 & neut.\ cont. & 100 \\
elec.\ cont. & 10 & energy & 1 & ion mom.\ $x$ & 20 \\
ion mom.\ $r$ & 20 & $n_n$ inlet & 200 & $V_{e\theta}$ inlet & 20 \\
$V_{ir}$ wall & 20 & $V_{er}$ wall & 10 & $J_T$ target & 100 \\
$J_T$ (A) & 5 & elec.\ mom.\ $x,r,\theta$ & $10^{-2}$ & data $n_i$ & 10 \\
data $T_e$ & 5 & data $V_p$ & 30 & smooth $n_i$ & $10^{-2}$ \\
smooth $n_n$ & $10^{-2}$ & smooth $\phi$ & $10^{-1}$ & smooth $T_e$ & $10^{-3}$ \\
\bottomrule
\end{tabular}
\end{table}

The weight hierarchy reflects the structure of the problem rather than ad hoc tuning. The continuity equations and the Poisson residual, which carry the global conservation structure and the electrostatic coupling, are weighted highest. The data terms sit an order of magnitude below, with the potential weighted most heavily since it is the field the probe data constrain most reliably. The electron momentum residuals carry $10^{-2}$, the smallest weight assigned to any governing equation, because $V_{ex}$ is fixed by the closure of Eq.~\eqref{eq:closure} and these equations act as transport closures rather than conservation statements; the consequence of that choice is examined in Sec.~\ref{sec:bohm}. The energy residual is weighted at unity because $T_e$ is directly anchored by the data term, so the energy equation shapes the temperature between data points rather than setting its overall level.

\subsection{Training procedure}

Training uses Adam \citep{kingma2014} at learning rate $10^{-4}$ with a ReduceLROnPlateau schedule (factor 0.5, patience 300, floor $10^{-6}$) for 15{,}000 epochs, from scratch, with no early stopping. The collocation points are drawn once at initialization and held fixed; resampling between epochs injected noise into the stiff Poisson residual and was not used. Gradients are clipped to unit norm to stabilize the early high-curvature phase. Two random seeds are fixed: the network initialization uses seed 7, and the collocation and boundary points are drawn with a separate seed (2601), so that both are reproducible and the initialization can be varied with the sampling held fixed.

We restrict training to first-order optimization. In an optimizer comparison carried out on a preliminary configuration of the model, adding an L-BFGS phase after Adam, a common refinement in PINN work, doubled the radial-banding diagnostics of the ion density, and one hybrid run converged to a discharge current of \SI{7.29}{\ampere} while the aggregate loss continued to fall. The SOAP optimizer \citep{vyas2024} reached a marginally lower data-fit loss but degraded every conservation residual, raising the ion continuity residual by a factor of four. All results reported here therefore use Adam alone.

\section{Experimental Data and Training}
\label{sec:data}

\subsection{The P5 internal probe dataset}

Validation data come from the internal plasma measurements performed at the University of Michigan Plasmadynamics and Electric Propulsion Laboratory on the P5 thruster \citep{haas2000,haas1999}. Using the High-speed Axial Reciprocating Probe (HARP) system, which limits probe residence time and thruster perturbation, Haas and Gallimore mapped the discharge channel with a double Langmuir probe, obtaining axial profiles of ion number density and electron temperature at five radial positions spanning the channel width at fixed \SI{300}{\volt} discharge voltage \citep{haas2000}; companion measurements provide the internal plasma potential \citep{haas1999}. At the \SI{5.4}{\ampere} (\SI{1.6}{\kilo\watt}) operating point used here, the reported maxima are an electron temperature of approximately \SI{38}{\electronvolt} and an ion density of approximately $2\times10^{18}\ \mathrm{m^{-3}}$, and the ion density exhibits a dual-peak axial structure indicating two ionization regions \citep{haas2000}, a feature that any two-dimensional model of this operating point should confront.

\subsection{Data reconstruction and anchoring}

Haas and Gallimore report their internal profiles graphically and provide no numerical data tables. The training targets were therefore reconstructed from the published figures rather than digitized point by point. For each of the five radial positions, the axial profiles of $n_i$, $T_e$, and $V_p$ were represented by smooth analytical curves reproducing the reported peak amplitudes, peak locations, dual-peak structure, and overall axial shape \citep{haas2000,haas1999}. This reconstruction introduces a modeling step between the original measurements and the training targets; we disclose it explicitly, as it bears on the interpretation of the reported fit errors, which measure agreement with the reconstructed profiles rather than with the raw probe data. Comparisons of shape along an axial sweep survive the reconstruction better than comparisons of absolute amplitude between radii, and the results below should be read with that asymmetry in mind.

\subsection{Measurement uncertainty}
\label{subsec:uncertainty}

The uncertainty of the original measurements is substantial and is reported by Haas \citep{haas2001thesis}. Ion number density and electron temperature, obtained from the double Langmuir probe through standard probe theory, carry uncertainties of 50\% and 20\%, respectively \citep{haas2001thesis}. The plasma potential, measured with the floating emissive probe, carries an asymmetric uncertainty of \SIrange{-3}{+6}{\volt} associated with the heater current that keeps the filament emitting \citep{haas2001thesis}, and the axial probe position is uncertain by $\pm\SI{1.5}{\milli\meter}$ because the probes were aligned by hand before high-speed actuation \citep{haas2001thesis}. These bands are shown with the data in Fig.~\ref{fig:validation}. They set the scale against which the fit errors of Sec.~\ref{sec:results} should be judged: an ion-density error of 20\% lies well inside a 50\% measurement band, so the comparison constrains the shape and location of the density profile more tightly than its absolute amplitude. The reconstruction step adds an uncertainty of its own, not included in these bands, which is why the fit errors are reported against the reconstructed targets and labelled as such throughout.

The reconstruction yields $355$ points per field, organized as five radial positions ($r - r_{\mathrm{in}} = 2.5, 7.5, 12.5, 17.5, \SI{22.5}{\milli\meter}$ from the inner wall, equivalently $\hat{r} = 1.645, 1.776, 1.908, 2.039, 2.171$) sampled at $71$ axial stations spanning the full probe range. The experimental probes extend to roughly \SI{80}{\milli\meter} from the anode, well beyond the \SI{38}{\milli\meter} discharge channel modeled here; restricting the targets to the computational domain ($\hat{x}\le 1$) retains $145$ points per field, 29 per radius. Each field enters the loss through its anchored term in Eq.~\eqref{eq:loss_data} with the weights $(w_{n_i},w_{T_e},w_{V_p}) = (10,5,30)$ applied uniformly across the five radii, $n_i$ in log-density. No other field is constrained by data: the neutral density, the five velocity components, and the charge separation $\eta$ are fixed by the physics residuals and boundary conditions alone. Because the five radial positions are fixed, the data constrain radial structure at five locations and nowhere between them; the held-out tests of Sec.~\ref{sec:heldout} exploit this directly.

\subsection{Reproducibility}

All results for the reference solution reported in Sec.~\ref{sec:results} come from a single training run under the configuration tabulated above: the normalization of Table~\ref{tab:norm}, the boundary conditions of Table~\ref{tab:bc}, the network and loss weights of Tables~\ref{tab:network} and~\ref{tab:weights}, $15{,}000$ Adam epochs from a fixed seed, and no early stopping. The held-out runs of Sec.~\ref{sec:heldout} use the same configuration with only the data term modified. Reported metrics are evaluated on a fixed $200\times80$ grid from the trained model at the end of training, so that they are deterministic and reproducible from the archived checkpoint.

\section{Results}
\label{sec:results}

\subsection{Baseline solution}

\begin{figure}[ht]
\centering
\begin{subfigure}[t]{0.3\textwidth}
\includegraphics[width=\textwidth]{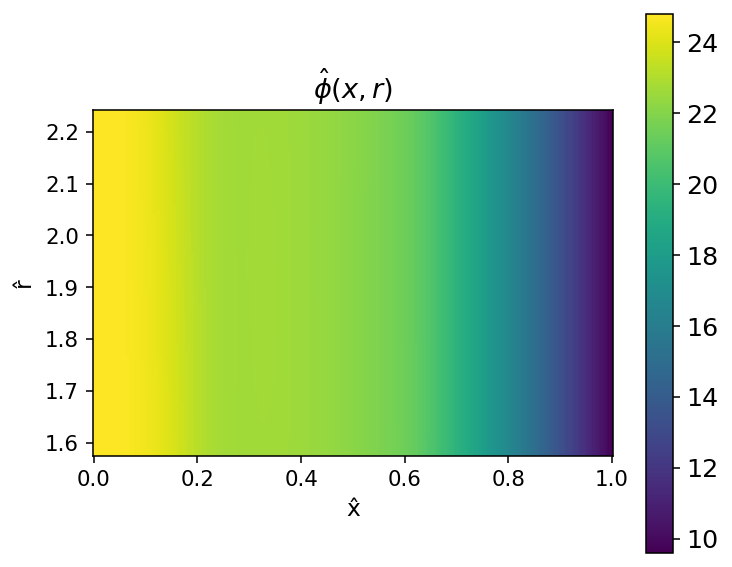}
\caption{Plasma potential $\hat{\phi}$.}
\end{subfigure}\hfill
\begin{subfigure}[t]{0.3\textwidth}
\includegraphics[width=\textwidth]{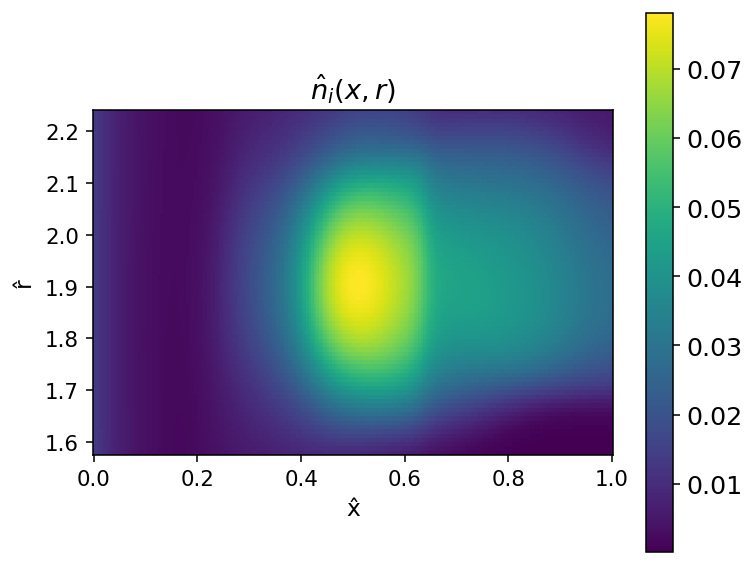}
\caption{Ion density $\hat{n}_i$.}
\end{subfigure}\hfill
\vspace{0em}
\begin{subfigure}[t]{0.3\textwidth}
\includegraphics[width=\textwidth]{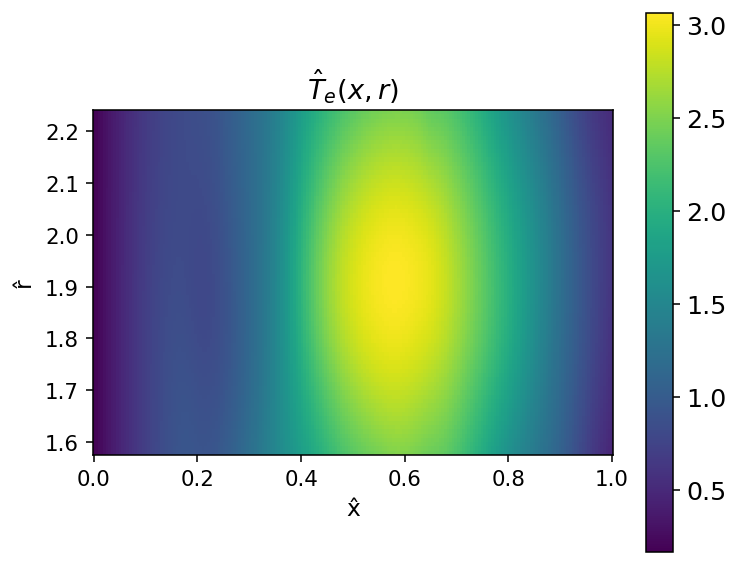}
\caption{Electron temperature $\hat{T}_e$.}
\end{subfigure}\hfill
\begin{subfigure}[t]{0.3\textwidth}
\includegraphics[width=\textwidth]{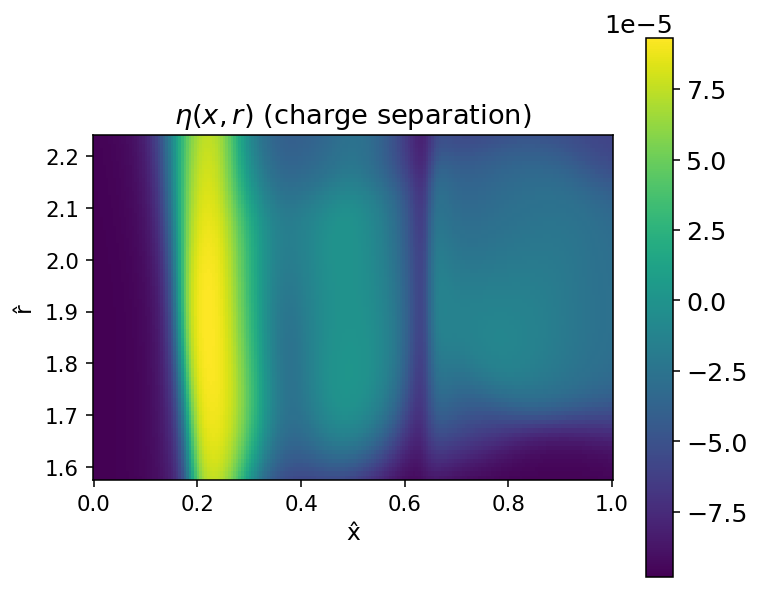}
\caption{Charge separation $\eta$.}
\end{subfigure}
\caption{Baseline solution: converged two-dimensional fields of the reference run over the full channel $(\hat{x},\hat{r})$, in the normalized units of Table~\ref{tab:norm}. The ion density peaks in the central channel where the rising temperature meets the neutral flow; $\eta$ shows its largest excursions across the upstream ionization layer and remains within $|\eta|\le10^{-4}$ everywhere.}
\label{fig:fields}
\end{figure}

Figure~\ref{fig:fields} shows four representative converged fields of the reference run over the full channel, $(\hat{x},\hat{r})\in[0,1]\times[\hat{r}_{\mathrm{in}},\hat{r}_{\mathrm{out}}]$: the electric potential, the ion density, the electron temperature, and the charge-separation field $\eta$. The potential decreases monotonically from \SI{300}{\volt} at the anode to \SI{116.2}{\volt} at the exit, so the axial field $E_x = -\partial_x\phi$ is positive throughout and steepens toward the exit plane where the magnetic barrier peaks. The ion density peaks in the central channel, where the rising electron temperature meets the not-yet-depleted neutral flow and the ionization source $S_i = k_i(T_e)\,n_n n_e$ is largest; downstream, the neutral density collapses and the source extinguishes. All densities and the electron temperature remain strictly positive. The charge separation spans $-9.8\times10^{-5}$ to $+9.3\times10^{-5}$, within its bound, with its largest excursions across the upstream ionization layer near $\hat{x}\approx0.2$, where the density gradients sourcing Poisson's equation are steepest; $|\eta|$ reaches 95\% of $\eta_{\max}$ over about 8\% of the domain, so the charge separation is neither negligible nor pinned against the constraint. The trained discharge current is $J_T = \SI{5.394}{\ampere}$, $-0.12\%$ from the \SI{5.4}{\ampere} target; the current is imposed through the penalties of Table~\ref{tab:weights} rather than predicted.

\subsection{Agreement with the Haas--Gallimore training profiles}

\begin{figure}[ht]
\centering
\begin{subfigure}[t]{0.49\textwidth}
\includegraphics[width=\textwidth]{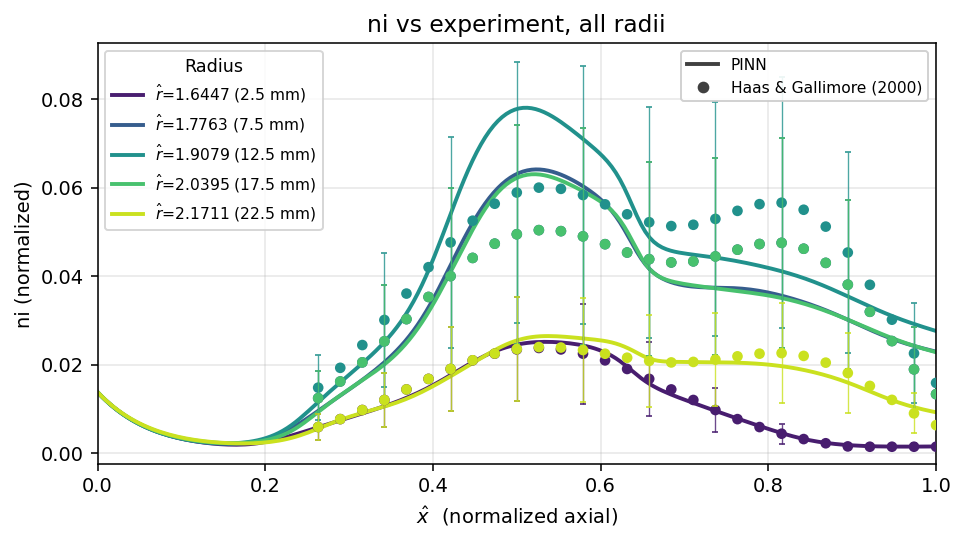}
\caption{Ion density.}
\end{subfigure}\hfill
\begin{subfigure}[t]{0.49\textwidth}
\includegraphics[width=\textwidth]{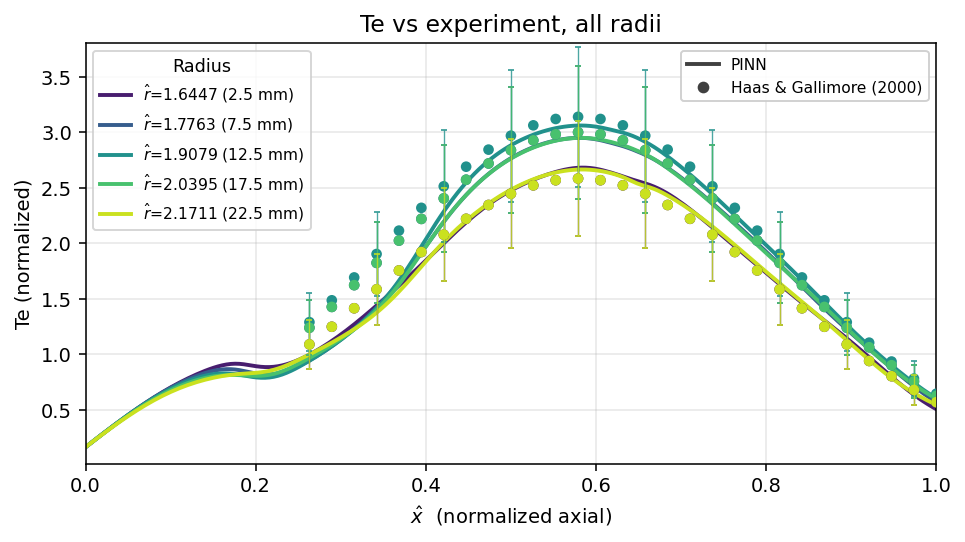}
\caption{Electron temperature.}
\end{subfigure}\\[4pt]
\begin{subfigure}[t]{0.49\textwidth}
\includegraphics[width=\textwidth]{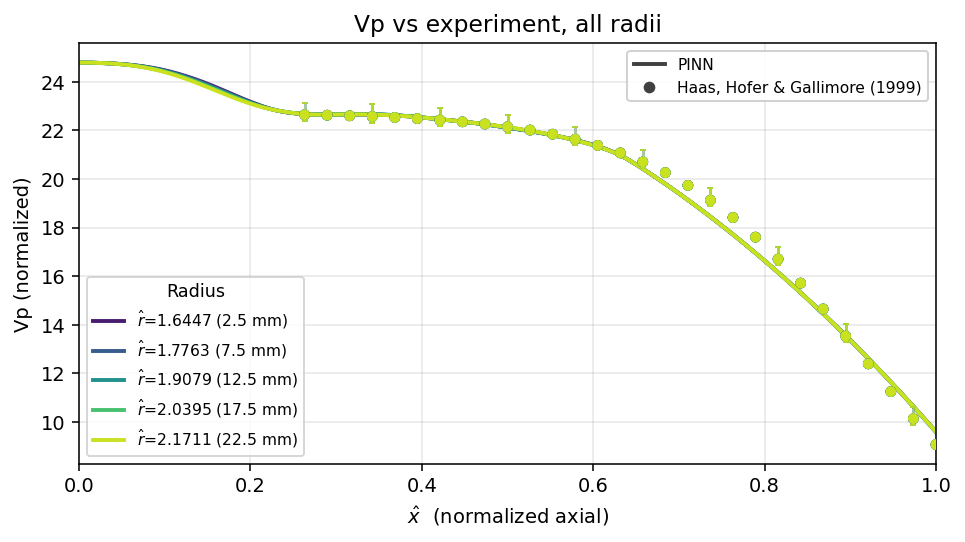}
\caption{Plasma potential.}
\end{subfigure}
\caption{PINN prediction (lines) against the reconstructed Haas--Gallimore profiles (symbols) at all five radial positions, labelled by their distance from the inner wall. Error bars show the measurement uncertainty reported by Haas \citep{haas2001thesis}: 50\% on $n_i$, 20\% on $T_e$, and \SIrange{-3}{+6}{\volt} on $V_p$. Amplitude comparisons between radii should be read with the reconstruction caveat of Sec.~\ref{sec:data}.}
\label{fig:validation}
\end{figure}

Table~\ref{tab:validation} summarizes the quantitative comparison with the reconstructed profiles over the 145 in-domain points per field, and Fig.~\ref{fig:validation} overlays the prediction on the data at all five radii. The plasma potential is reproduced with a relative $L^2$ error of $0.017$ and the electron temperature with $0.073$; both the amplitude (\SI{37.05}{\electronvolt} vs.\ ${\sim}\SI{38}{\electronvolt}$) and the location ($\hat{x} = 0.583$ vs.\ ${\approx}0.55$) of the temperature peak are captured. The ion density is the most demanding field, with a relative $L^2$ error of $0.205$. Its peak location on the centerline ($\hat{x} = 0.513$ vs.\ ${\approx}0.53$) matches the measurements, but its amplitude, normalized $0.078$ or $2.62\times10^{18}\ \mathrm{m^{-3}}$, exceeds the measured ${\sim}2\times10^{18}\ \mathrm{m^{-3}}$ by about 30\%, an overestimate that remains inside the 50\% uncertainty of the density measurement (Sec.~\ref{subsec:uncertainty}). The electron-temperature prediction lies within its measurement band at all five radii (Fig.~\ref{fig:validation}). The ion-density prediction does so over most of the channel, including the peak, but not near the exit, where it exceeds the reconstructed data by 46 to 74\% at four of the five radii. The potential error of $0.017$ corresponds to about \SI{4}{\volt} RMS over the \SIrange{116}{300}{\volt} range, comparable to the \SIrange{-3}{+6}{\volt} emissive-probe uncertainty; the prediction lies within that band over most of the channel but falls a few volts below it locally, in the acceleration region near $\hat{x}\approx0.65$--$0.8$, a deficit that Sec.~\ref{subsec:phibound} traces to the bound of the potential representation. The two-dimensional model reproduces the ordering of the profiles across the channel, with the inner- and outer-wall radii carrying roughly a third of the central density. The downstream ionization peak reported by Haas and Gallimore near $\hat{x}\approx0.8$ is not reproduced as a separate maximum; the prediction carries it only as a shoulder in the decay of the density beyond the main peak (Fig.~\ref{fig:validation}a).

\begin{figure}[ht]
\centering
\includegraphics[width=\textwidth,trim={0 0 0 45bp},clip]{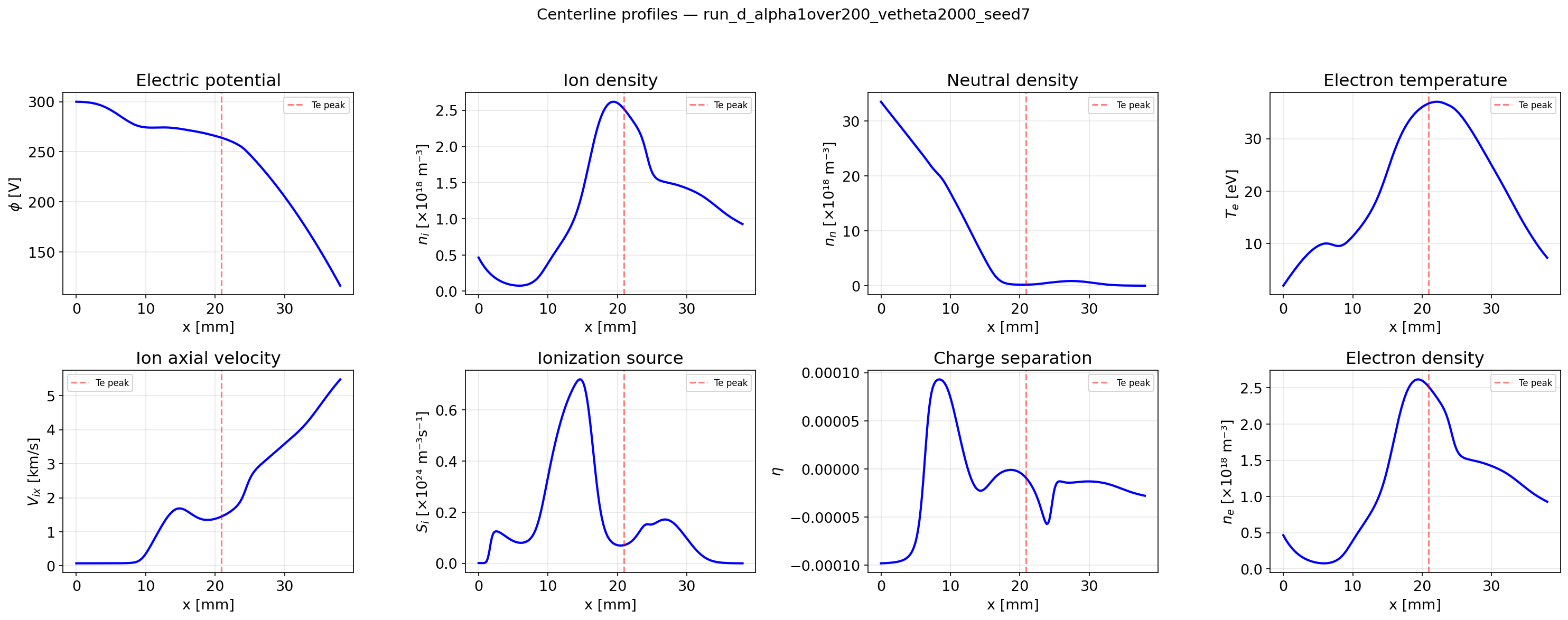}
\caption{Centerline profiles of the reference run in dimensional units: potential, ion, neutral and electron densities, electron temperature, ion axial velocity, ionization source, and charge-separation field $\eta$.}
\label{fig:centerline}
\end{figure}

The centerline exit ion velocity is \SI{5.48}{\kilo\meter\per\second}, about half the \SI{11}{\kilo\meter\per\second} measured by laser-induced fluorescence \SI{1}{\milli\meter} downstream of the exit plane of the P5 at the neighboring condition of \SI{300}{\volt}, \SI{5.3}{\ampere} \citep{williams1999}; the same measurements give \SI{16}{\kilo\meter\per\second} at \SI{100}{\milli\meter}, the ions continuing to accelerate outside the channel. That corresponds to roughly \SI{20}{\volt} of acceleration against the \SI{184}{\volt} dropped across the modeled domain, so the ions leave with about 11\% of the kinetic energy the modeled potential drop would give them. A plausible principal contributor is the ion--neutral friction at the prescribed constant rate $\nu_{in} = \SI{1.8e6}{\per\second}$, a limitation of the reduced fluid closure rather than of the PINN solution of it. In normalized form this rate is $\hat{\nu}_{in} = \nu_{in}\tau^{*} \approx 23$, so an ion undergoes some twenty collisions during the reference transit time $\tau^{*}$, and at \SI{15}{\kilo\meter\per\second} its stopping length $V_{ix}/\nu_{in}$ is about \SI{8}{\milli\meter}, a fifth of the channel. Ions therefore cannot approach free-fall acceleration: the axial momentum balance of Eq.~\eqref{eq:ionmomx} drives them instead toward a drag-limited velocity at which the electric force is balanced by the friction term, $V_{ix} - V_n \approx (\partial\phi/\partial x)/\hat{\nu}_{in}$ when $m_n \simeq m_i$, and that velocity falls short of the free-fall value over the modeled domain. Because the rate is held constant, the friction also persists in the downstream region where the neutral density has collapsed (Fig.~\ref{fig:centerline}), whereas a rate evaluated from the local neutral density would vanish there. Two further features of the model act in the same direction but more weakly: ionization loading, which injects new ions at the neutral velocity, and the absence of ion pressure. The shortfall is the largest single discrepancy between the model and the physical device, and the closure that produces it is identified rather than compensated: no weight was adjusted to recover the exit velocity. Section~\ref{sec:consistency} shows that once the rate is evaluated locally and the global balances are corrected, the exit velocity is limited instead by the tension between the measured density and the measured potential drop.

\begin{table}[ht]
\centering
\caption{Validation of the reference run against the reconstructed Haas--Gallimore data \citep{haas2000,haas1999} (relative $L^2$ over the 145 in-domain points per field), with field metrics of the converged solution. The right column gives the measurement uncertainty reported by Haas \citep{haas2001thesis} for the three fitted fields. The discharge current is imposed through penalties and is listed for completeness.}
\label{tab:validation}
\begin{tabular}{lll}
\toprule
Quantity & Model & Haas--Gallimore \\
\midrule
$V_p$ fit ($L^2$ rel.) & 0.0173 & uncertainty \SIrange{-3}{+6}{\volt} \\
$T_e$ fit ($L^2$ rel.) & 0.0730 & uncertainty 20\% \\
$n_i$ fit ($L^2$ rel.) & 0.2048 & uncertainty 50\% \\
$n_i$ peak (centerline) & $0.078\;/\;2.62\times10^{18}\ \mathrm{m^{-3}}$ & ${\sim}2.0\times10^{18}\ \mathrm{m^{-3}}$ \\
$n_i$ peak location & $\hat{x} = 0.513$ & $\hat{x} \approx 0.53$ \\
$T_e$ peak & \SI{37.05}{\electronvolt} & ${\sim}\SI{38}{\electronvolt}$ \\
$T_e$ peak location & $\hat{x} = 0.583$ & $\hat{x} \approx 0.55$ \\
$\eta$ range & $-9.8\times10^{-5}$ to $+9.3\times10^{-5}$ & (bound $10^{-4}$) \\
$V_{ix}$ exit (centerline) & \SI{5.48}{\kilo\meter\per\second} & ${\approx}\SI{11}{\kilo\meter\per\second}$ (LIF \citep{williams1999}) \\
$J_T$ & \SI{5.394}{\ampere} & \SI{5.4}{\ampere} (target) \\
\bottomrule
\end{tabular}
\end{table}

Figure~\ref{fig:centerline} collects the centerline profiles of the principal fields in dimensional units. The neutral density depletes by ionization over the first half of the channel; the ionization source forms a localized zone upstream of the temperature peak; and the charge-separation field $\eta$ exhibits its largest excursions across the ionization layer while remaining within its prescribed bound.

\FloatBarrier

\subsection{Comparison with a published hybrid particle-in-cell simulation}
\label{subsec:koo}

The P5 internal dataset has also served as the validation target for conventional simulation. Koo and Boyd simulated the P5 with a two-dimensional axisymmetric hybrid particle-in-cell/Monte Carlo collision (PIC-MCC) code, with kinetic heavy species and a fluid electron model \citep{koo2001,koo2005}, and compared the results against the same Haas measurements used here. Table~\ref{tab:koo} sets their reported agreement alongside that of the present model. Two differences must be stated before the comparison is read. First, the operating points differ: the hybrid-PIC results were obtained at \SI{3}{\kilo\watt} (\SI{300}{\volt}, \SI{10}{\ampere}) \citep{koo2005}, and no published simulation of the P5 at the \SI{1.6}{\kilo\watt} condition modeled here was found, so the comparison is between levels of agreement with experiment rather than point by point. Second, the two approaches use the measurements differently. The hybrid-PIC model is a forward solver into which the data enter only through a semi-empirical electron mobility constructed in part from the measurements \citep{koo2005}; the PINN assimilates the measured $n_i$, $T_e$, and $V_p$ directly as loss terms. The PINN errors of Table~\ref{tab:koo} are therefore fit errors and are not a statement of superior predictive accuracy.

\begin{table}[ht]
\centering
\caption{Agreement with the Haas measurements \citep{haas2001thesis} reported for the hybrid PIC-MCC simulation of Koo and Boyd \citep{koo2005} and obtained for the present PINN. The two models are compared at different operating points and use the data differently (see text); the comparison is between levels of agreement, not a point-by-point benchmark.}
\label{tab:koo}
\setlength{\tabcolsep}{6pt}
\begin{tabular}{lll}
\toprule
 & Hybrid PIC-MCC \citep{koo2005} & Present PINN \\
\midrule
Operating point & \SI{300}{\volt}, \SI{10}{\ampere} (\SI{3}{\kilo\watt}) & \SI{300}{\volt}, \SI{5.4}{\ampere} (\SI{1.6}{\kilo\watt}) \\
Use of data & semi-empirical mobility & loss terms on $n_i$, $T_e$, $V_p$ \\
$V_p$ & centerline profile closely replicated & $L^2 = 0.017$, five radii \\
$n_i$ & within experimental uncertainty & $L^2 = 0.205$, mostly within 50\% band \\
$T_e$ & off by a factor of two; incorrect & $L^2 = 0.073$; energy residual \\
 & near-field structure & RMS $0.499$ \\
\bottomrule
\end{tabular}
\end{table}

Read with those qualifications, the comparison is informative in one respect. Both models reproduce the potential well and the ion density largely within the measurement uncertainty, and both find the electron temperature to be the field the reduced physics handles worst. In the hybrid-PIC simulation the temperature is off by a factor of two in magnitude and has a different near-field structure, and Koo notes that the electron temperatures produced by that code had shown ``extremely weak correlation with experimental data'' \citep{koo2005}. In the PINN the same difficulty appears not as a temperature error, since $T_e$ is held near the data by its loss term, but as the largest equation residual at convergence (Table~\ref{tab:residuals}): the electron energy equation cannot be satisfied simultaneously with the measured temperature. The two formulations thus locate the same weakness of the fluid electron energy model from opposite sides, the forward solver through its prediction and the PINN through its residual. The PINN supplies a converged two-dimensional field set consistent with the measurements to within their uncertainty, and it reports the equation residual that measures how far the reduced physics departs from them.

\subsection{Convergence}

\begin{figure}[ht]
\centering
\begin{subfigure}[t]{0.36\textwidth}
\includegraphics[width=\textwidth]{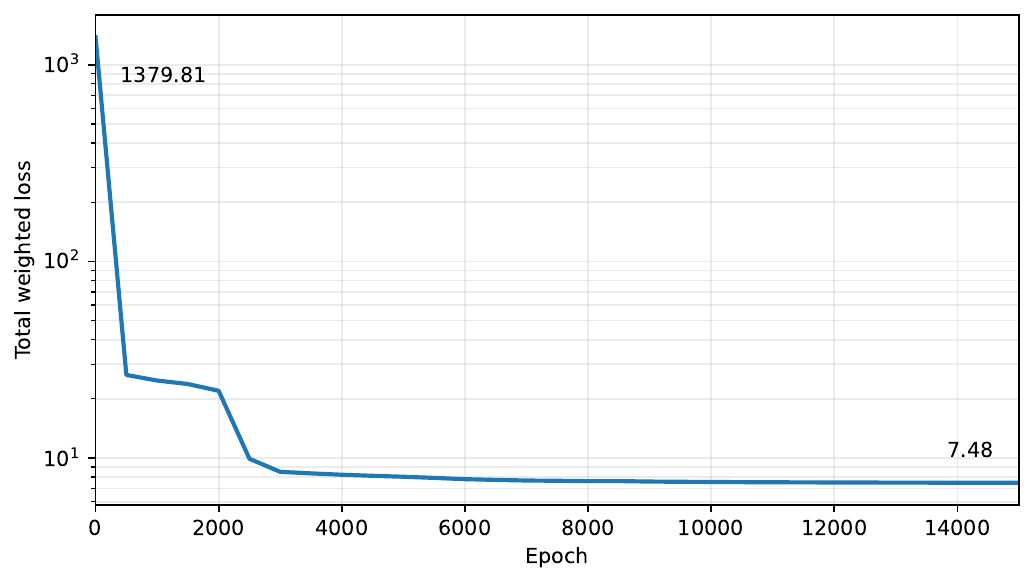}
\caption{Total weighted loss.}
\end{subfigure}\hfill
\begin{subfigure}[t]{0.62\textwidth}
\includegraphics[width=\textwidth]{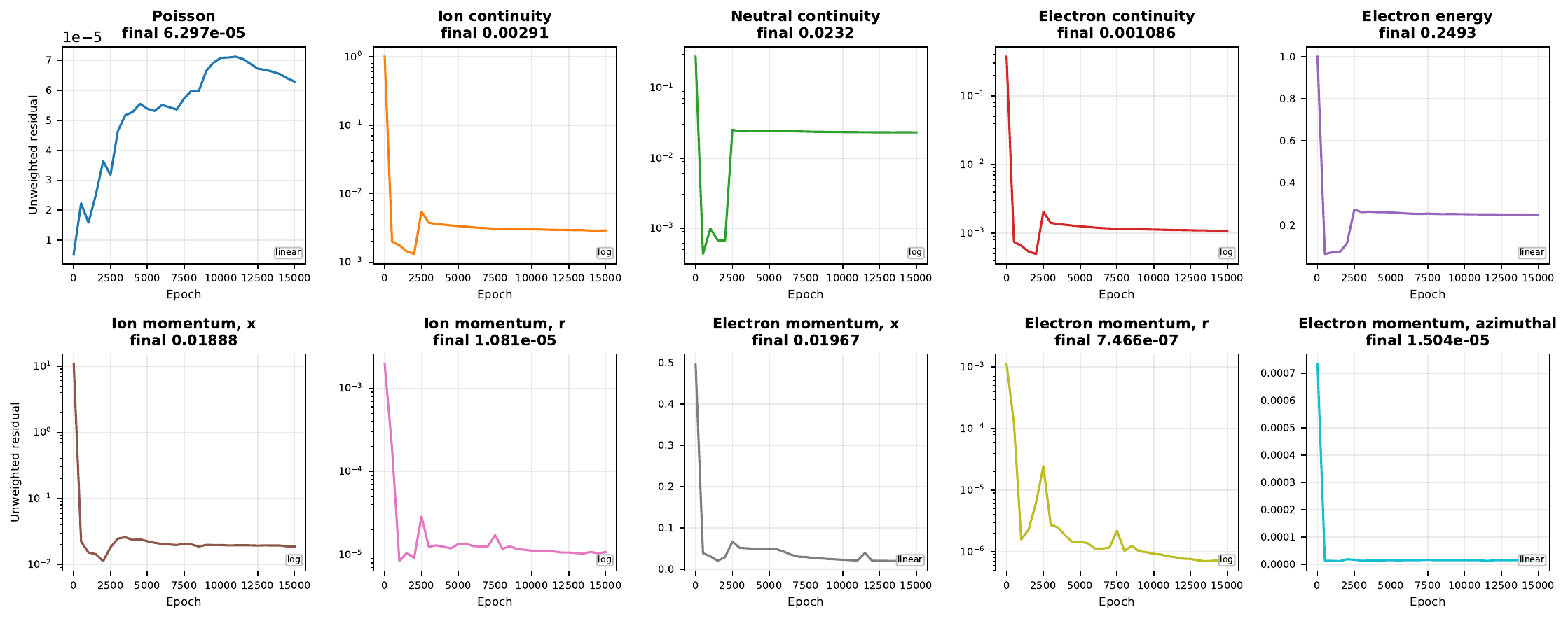}
\caption{Unweighted residual of each governing equation.}
\end{subfigure}
\caption{Training history of the reference run over 15{,}000 Adam epochs. The per-equation panels are autoscaled individually, since the residuals span nearly six decades at convergence; the values they print are those of the training collocation set, while Table~\ref{tab:residuals} reports the evaluation grid.}
\label{fig:convergence}
\end{figure}

Figure~\ref{fig:convergence}a shows the training history. The total weighted loss decreases smoothly from $1379.81$ at initialization to $7.48$ after 15{,}000 Adam epochs, with most of the reduction occurring within the first few hundred epochs. Over the final thousand epochs, the loss changes by only $9.3\times10^{-3}$, indicating that training has reached a near-stationary regime. Convergence is assessed more directly through the stabilization of the individual governing-equation residuals, reported in Fig.~\ref{fig:convergence}b and Table~\ref{tab:residuals}.

In the classification of Sec.~\ref{subsec:regimes}, a purely unsupervised forward problem \citep{burrows2024} can drive its equation residuals arbitrarily low: the residuals are the only quantities the optimization minimizes, and nothing in the loss constrains the solution toward measurement. A formulation that assimilates experimental data instead carries a residual floor set by the mismatch between the reduced fluid description and the measurements, so residual magnitudes are not comparable across the two regimes. The weighted total, moreover, sums heterogeneous terms under weights spanning several orders of magnitude and has no absolute scale: it can fall by improving a heavily weighted term while a lightly weighted one is left untouched. The convergence metric reported here is therefore the unweighted residual of each governing equation (Fig.~\ref{fig:convergence}b and Table~\ref{tab:residuals}).

The continuity and momentum residuals are reduced by one to three orders of magnitude from initialization. The largest residual at convergence is the electron energy equation, with a mean squared value of $0.249$ (RMS $0.499$) out of a total of $0.316$. It settles early in training and does not decrease thereafter, while the temperature data term continues to fall. We interpret this as a modeling limitation of the energy balance rather than an optimization failure: the equation carries convection, Ohmic heating, collisional exchange, and ionization but no conductive heat flux and no wall losses, and with these terms alone it cannot reproduce the measured temperature profile. The data term, weighted above the energy residual, holds $T_e$ close to the measurements, and the residual records the extent to which the reduced energy equation disagrees with them. The same tension appears in conventional simulations of this thruster, where the electron temperature is the field least well predicted (Sec.~\ref{subsec:koo}). The rescaled Poisson residual rises during training, from $5.3\times10^{-6}$ to $6.3\times10^{-5}$; this reflects the initialization rather than a degradation, since the run starts with $\eta$ at zero and $\phi$ close to its linear baseline, so both terms of the residual begin near zero for reasons unrelated to accuracy, and the growth records the solution becoming physical.

\begin{table}[ht]
\centering
\caption{Per-equation unweighted mean squared residuals of the reference run at initialization and at the end of training, with the final RMS, evaluated on the fixed $200\times80$ grid. Residuals marked ``scaled'' or ``rescaled'' carry the factors of Sec.~\ref{sec:loss}.}
\label{tab:residuals}
\begin{tabular}{lccc}
\toprule
Equation & Initial MSE & Final MSE & Final RMS \\
\midrule
Poisson (rescaled) & $5.34\times10^{-6}$ & $6.31\times10^{-5}$ & $7.94\times10^{-3}$ \\
Ion continuity (scaled) & $1.015$ & $2.87\times10^{-3}$ & $0.0536$ \\
Neutral continuity & $0.277$ & $2.32\times10^{-2}$ & $0.152$ \\
Electron continuity & $0.378$ & $1.07\times10^{-3}$ & $0.0327$ \\
Electron energy (scaled) & $1.035$ & $0.249$ & $0.499$ \\
Ion momentum $x$ (scaled) & $10.56$ & $1.84\times10^{-2}$ & $0.136$ \\
Ion momentum $r$ (scaled) & $1.91\times10^{-3}$ & $1.08\times10^{-5}$ & $3.28\times10^{-3}$ \\
Electron momentum $x$ & $0.485$ & $2.20\times10^{-2}$ & $0.148$ \\
Electron momentum $r$ & $1.10\times10^{-3}$ & $7.44\times10^{-7}$ & $8.62\times10^{-4}$ \\
Electron momentum $\theta$ & $7.23\times10^{-4}$ & $1.52\times10^{-5}$ & $3.90\times10^{-3}$ \\
\bottomrule
\end{tabular}
\end{table}

\FloatBarrier
\subsection{Training cost and seed sensitivity}
\label{subsec:cost}

A training run of 15{,}000 Adam epochs takes about \SI{4}{\hour} of wall-clock time on a laptop CPU (Apple M4) using four threads in double precision, about \SI{1}{\second} per epoch for the $1.2\times10^{4}$ collocation points of Table~\ref{tab:network}; no GPU is required. The cost is dominated by the second derivatives required by Poisson's equation, evaluated through automatic differentiation at every collocation point. Once trained, the network evaluates the eleven fields on the $200\times80$ evaluation grid in about \SI{28}{\milli\second}, and the full set of derived quantities, including the electric field components obtained by automatic differentiation, in about \SI{42}{\milli\second}. Because the training is a single optimization from scratch, each held-out configuration of Sec.~\ref{sec:heldout} carries the full training cost.

To assess sensitivity to the random initialization, the reference configuration was retrained with two further initialization seeds (1234 and 2601, in addition to the reference seed 7), the collocation and boundary points and all other settings held fixed. Across the three seeds the relative $L^2$ errors range over $0.2048$--$0.2070$ for $n_i$, $0.0718$--$0.0730$ for $T_e$, and $0.01730$--$0.01732$ for $V_p$; the trained discharge current varies by less than \SI{0.1}{\milli\ampere} about \SI{5.394}{\ampere}, the centerline exit ion velocity lies between $5.42$ and \SI{5.48}{\kilo\meter\per\second}, the electron energy residual RMS between $0.498$ and $0.499$, and the final total loss between $7.477$ and $7.484$. The seed-to-seed spread is at most about 1.6\% of any reported error, an order of magnitude below the differences discussed in this paper, so the results are insensitive to initialization.

\subsection{Effect of the anomalous transport term}
\label{sec:bohm}

The electron momentum block of Eqs.~\eqref{eq:emomx}--\eqref{eq:emomtheta} is the part of the model least directly constrained by the data: none of its unknowns is measured, and it carries the smallest weights in the objective. To assess how much the measured fields depend on it, the reference run is compared with a classical-transport configuration trained at the same seed, collocation set, calibration scales, and loss weights. That configuration omits the anomalous term, so that $\nu = \nu_{ei}+\nu_{en}$, and bounds the azimuthal velocity at $|V_{e\theta}|\le0.0125$ (about \SI{37}{\meter\per\second}) instead of 2000. The two differences were not varied independently, so the comparison characterizes the change of transport configuration as a whole rather than isolating the Bohm coefficient.

The Hall parameter responds as expected. Averaged over the evaluation grid, $\Omega$ falls from 6696 under classical transport, an order of magnitude above the range usually reported for these thrusters, to 119 in the reference run. Wherever $\omega_{ce}$ dominates the classical rates, the anomalous term caps the Hall parameter at $\Omega = \omega_{ce}/(\nu_c + \alpha_B\omega_{ce}) \to 1/\alpha_B = 200$, which is the value it takes over most of the downstream half of the channel. The measured fields, by contrast, barely move: the relative $L^2$ errors of the two configurations agree to within 2\% on $n_i$ and $V_p$ and 4\% on $T_e$, and the radial variation of the ion density at mid-channel changes only from 84.0\% to 87.7\% of the measured variation. A mean Hall parameter changing by a factor of 56 therefore produces data fits that are practically indistinguishable, and the electron momentum block is weakly coupled to the fields the objective measures. This is a statement about the formulation rather than about either run: a large part of the system can change substantially without the measured solution responding, so the aggregate loss and the data errors are poor guides to whether the electron transport is right.

\begin{figure}[ht]
\centering
\begin{subfigure}[t]{0.48\textwidth}
\includegraphics[width=\textwidth]{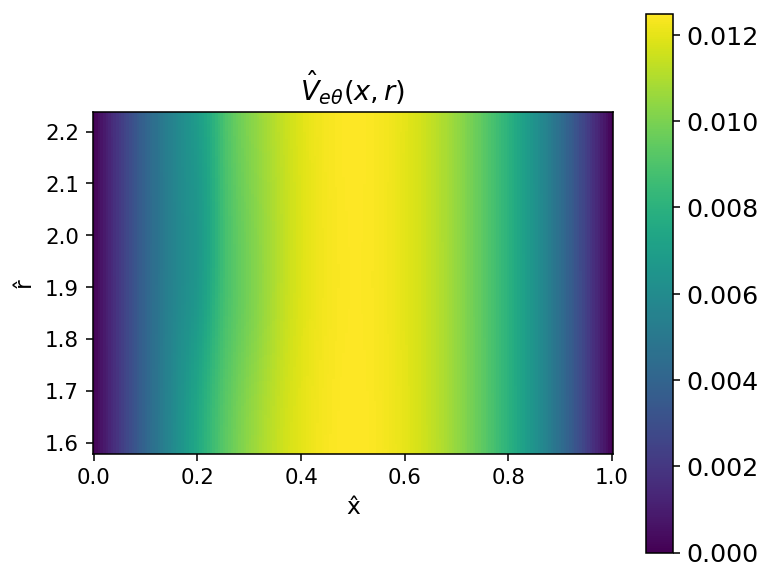}
\caption{Without the anomalous term (classical transport).}
\end{subfigure}\hfill
\begin{subfigure}[t]{0.48\textwidth}
\includegraphics[width=\textwidth]{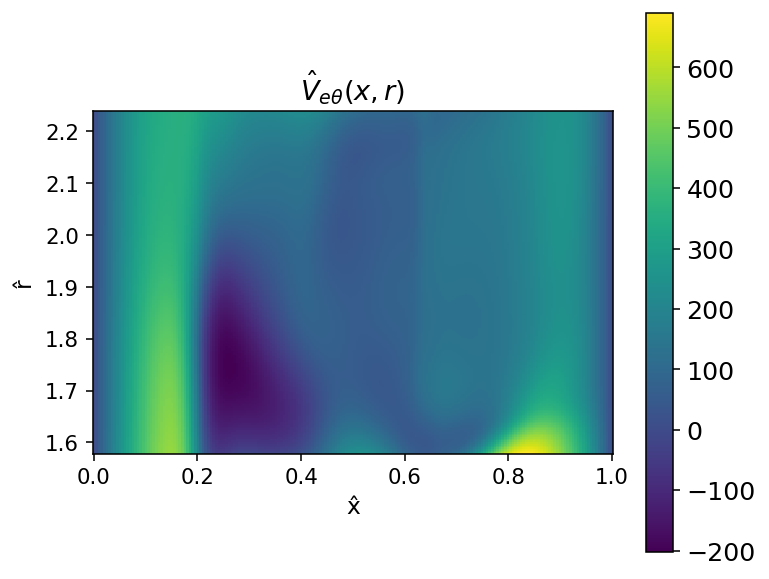}
\caption{With the anomalous term (reference run).}
\end{subfigure}
\caption{Azimuthal electron velocity $\hat{V}_{e\theta}$ without and with the anomalous transport term, in normalized units. Note the different color scales, which differ by more than four orders of magnitude: the classical-transport solution is bounded at $0.0125$ and follows the $\xi(1-\xi)$ profile of its envelope, whereas the reference solution reaches 689 with a two-dimensional structure set by the governing equations.}
\label{fig:vetheta}
\end{figure}

The azimuthal velocity shows the same decoupling most directly (Fig.~\ref{fig:vetheta}). Under classical transport, $V_{e\theta}$ is set by its envelope rather than by Eq.~\eqref{eq:emomtheta}: the solution presses against 99.96\% of the permitted range at its maximum and exceeds 95\% of the local envelope over about 72\% of the domain, so its spatial structure is the $\xi(1-\xi)$ profile of the parametrization. In the reference run, the converged $|V_{e\theta}|$ reaches 689 in normalized units, about \SI{2.1e6}{\meter\per\second}, which is 34\% of the permitted range. It develops a genuine two-dimensional structure, with a reversal region near the inner wall around $\hat{x}\approx0.25$ and a maximum near the inner wall at the exit, and it stays below 95\% of the local envelope everywhere. In that configuration the azimuthal momentum equation is satisfied to a mean squared residual of $1.52\times10^{-5}$ (Table~\ref{tab:residuals}), yet the four orders of magnitude separating the two azimuthal velocities leave the measured fields within a few percent of each other.

\FloatBarrier

\subsection{Complete field set of the reference solution}
\label{sec:allfields}

\begin{figure}[p]
\centering
\begin{subfigure}[t]{0.48\textwidth}
\includegraphics[width=\textwidth]{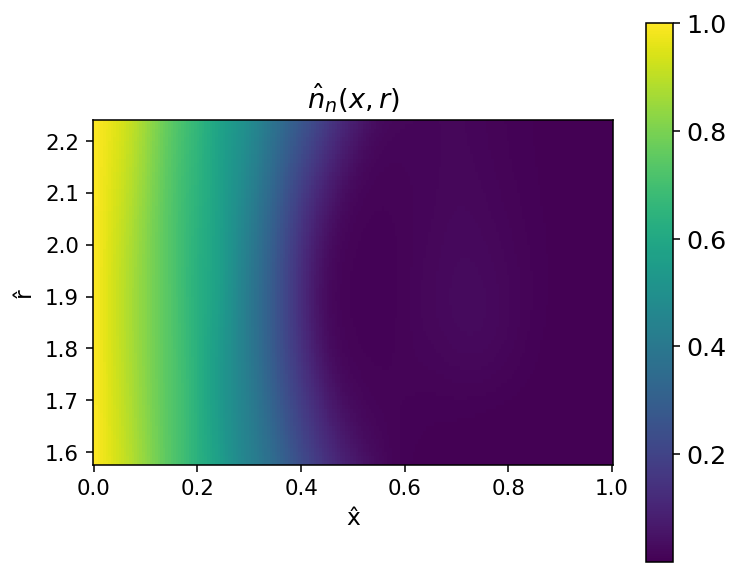}
\caption{Neutral density $\hat{n}_n$.}
\end{subfigure}\hfill
\begin{subfigure}[t]{0.48\textwidth}
\includegraphics[width=\textwidth]{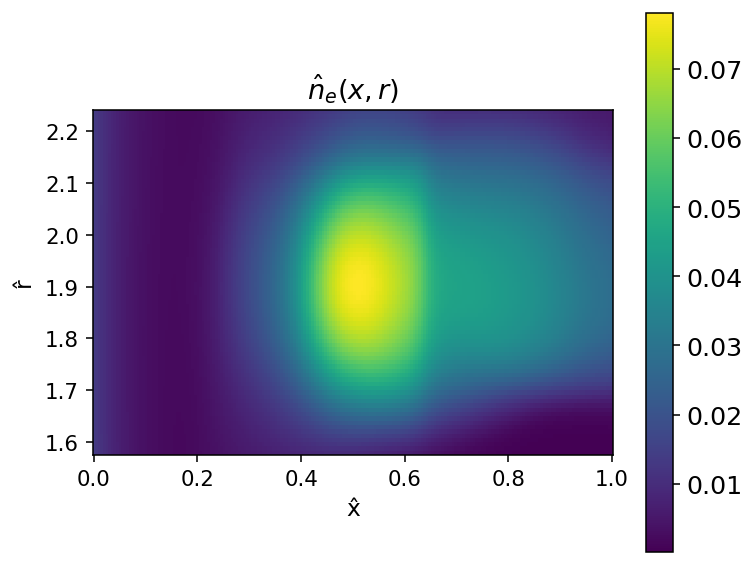}
\caption{Electron density $\hat{n}_e$.}
\end{subfigure}\\[4pt]
\begin{subfigure}[t]{0.48\textwidth}
\includegraphics[width=\textwidth]{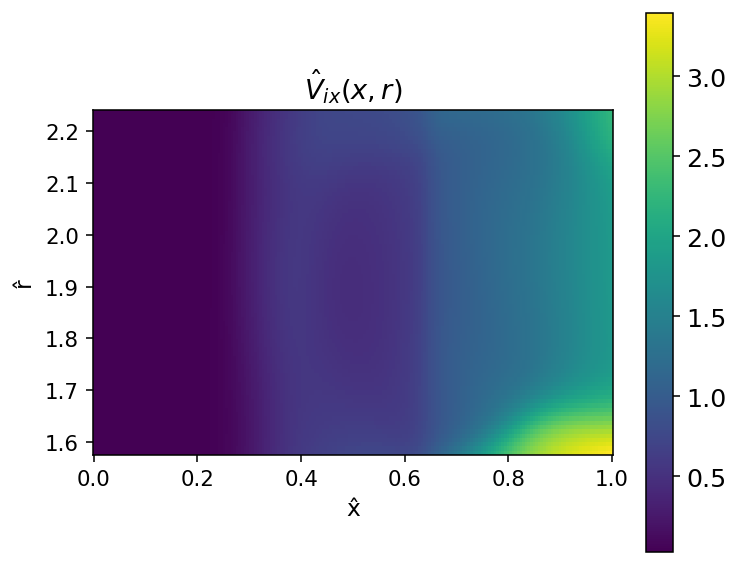}
\caption{Axial ion velocity $\hat{V}_{ix}$.}
\end{subfigure}\hfill
\begin{subfigure}[t]{0.48\textwidth}
\includegraphics[width=\textwidth]{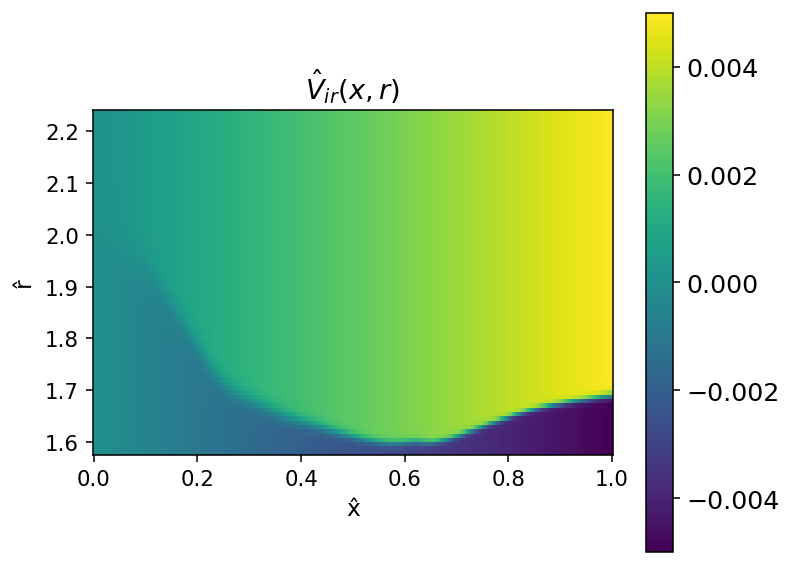}
\caption{Radial ion velocity $\hat{V}_{ir}$.}
\end{subfigure}\\[4pt]
\begin{subfigure}[t]{0.48\textwidth}
\includegraphics[width=\textwidth]{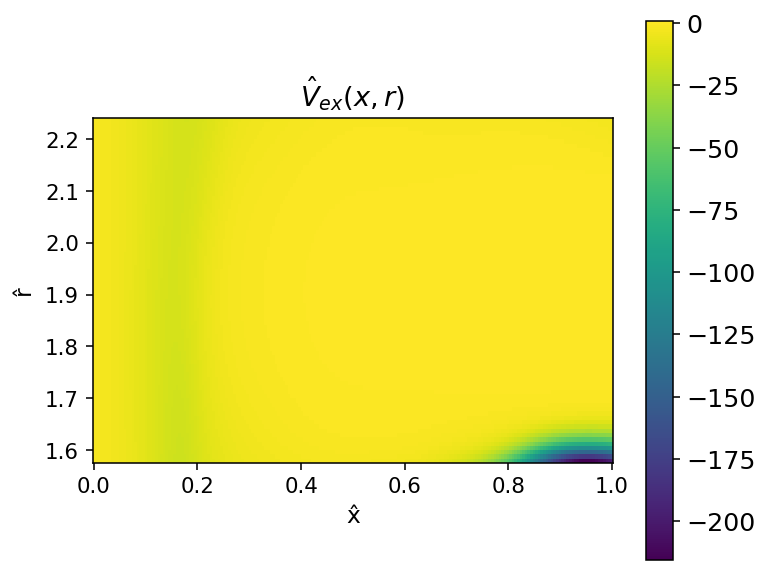}
\caption{Axial electron velocity $\hat{V}_{ex}$.}
\end{subfigure}\hfill
\begin{subfigure}[t]{0.48\textwidth}
\includegraphics[width=\textwidth]{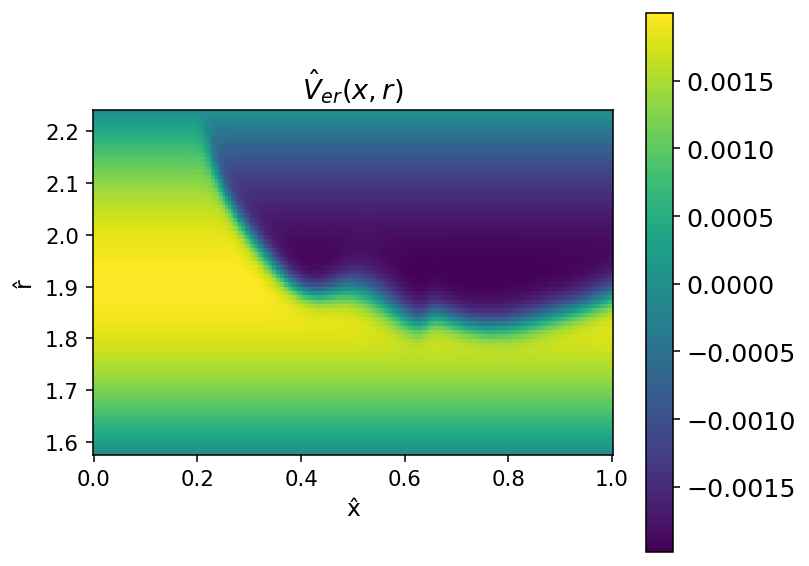}
\caption{Radial electron velocity $\hat{V}_{er}$.}
\end{subfigure}
\caption{Densities and velocities of the reference run, in normalized units. The azimuthal electron velocity is shown in Fig.~\ref{fig:vetheta}.}
\label{fig:all_densvel}
\end{figure}

\begin{figure}[p]
\centering
\begin{subfigure}[t]{0.48\textwidth}
\includegraphics[width=\textwidth]{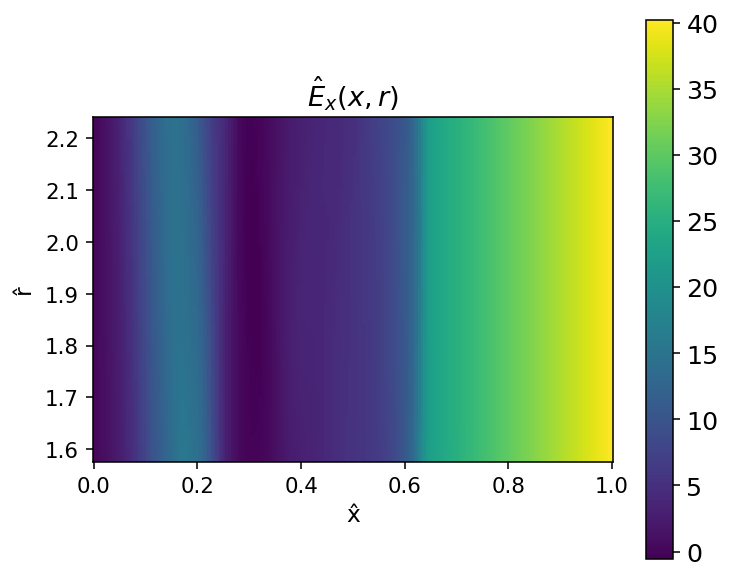}
\caption{Axial electric field $\hat{E}_x$.}
\end{subfigure}\hfill
\begin{subfigure}[t]{0.48\textwidth}
\includegraphics[width=\textwidth]{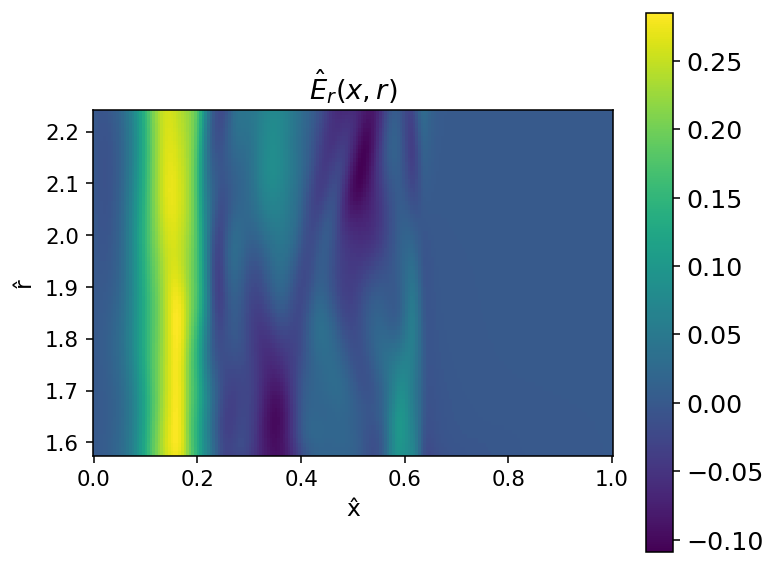}
\caption{Radial electric field $\hat{E}_r$.}
\end{subfigure}\\[4pt]
\begin{subfigure}[t]{0.48\textwidth}
\includegraphics[width=\textwidth]{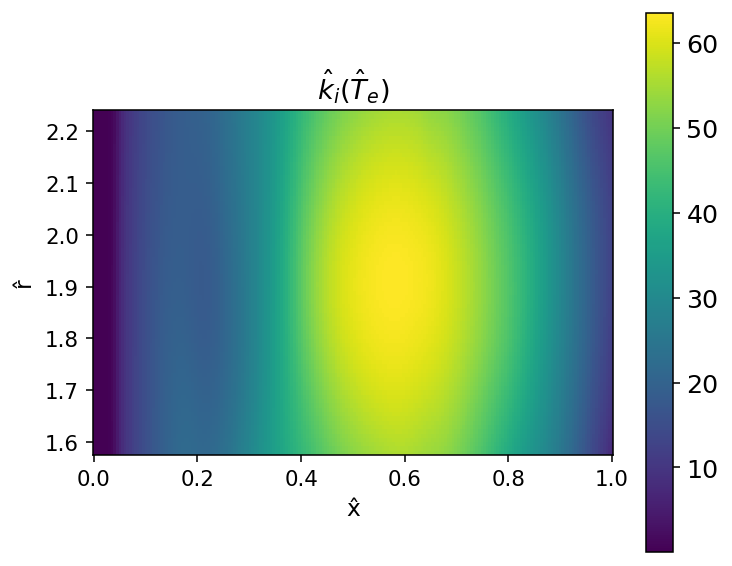}
\caption{Ionization rate coefficient $\hat{k}_i(T_e)$.}
\end{subfigure}\hfill
\begin{subfigure}[t]{0.48\textwidth}
\includegraphics[width=\textwidth]{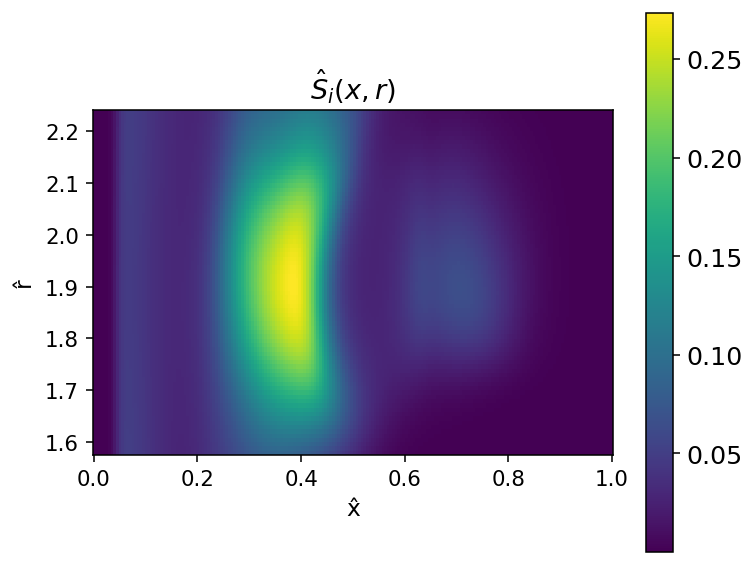}
\caption{Ionization source $\hat{S}_i$.}
\end{subfigure}
\caption{Electric field components and ionization quantities of the reference run, in normalized units.}
\label{fig:all_elec}
\end{figure}

For completeness, Figs.~\ref{fig:all_densvel} and~\ref{fig:all_elec} present the remaining fields of the reference run. The electron density (Fig.~\ref{fig:all_densvel}b) is indistinguishable from the ion density at plotting resolution, as expected for $|\eta|\le10^{-4}$. The axial ion velocity accelerates mainly in the last third of the channel, where the axial electric field (Fig.~\ref{fig:all_elec}a) steepens under the magnetic barrier; the electron axial velocity is negative throughout, as required for an anode-directed electron current, and is set by the current closure. The radial ion and electron velocities are small everywhere, bounded by their envelopes at $|V_{ir}|<0.005$ and $|V_{er}|\le0.002$. The radial electric field (Fig.~\ref{fig:all_elec}b) is retained in Eq.~\eqref{eq:emomr} rather than assumed to vanish; the banding upstream of $\hat{x}\approx0.65$ and the flat region beyond it are numerical and are not accounted for by the model. The ionization rate coefficient follows the electron temperature through Eq.~\eqref{eq:ki}, while the ionization source is localized upstream of the temperature peak, where the rate coefficient and the not-yet-depleted neutral density overlap.

\FloatBarrier

\subsection{Held-out tests}
\label{sec:heldout}

\begin{figure}[!htbp]
\centering
\begin{subfigure}[t]{0.49\textwidth}
\includegraphics[width=\textwidth]{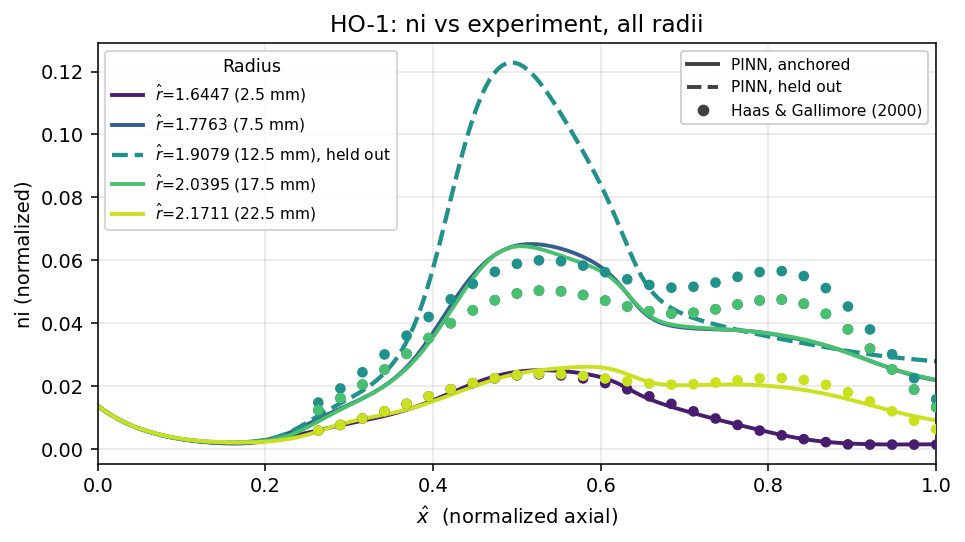}
\caption{HO-1: \SI{12.5}{\milli\meter} withdrawn.}
\end{subfigure}\hfill
\begin{subfigure}[t]{0.49\textwidth}
\includegraphics[width=\textwidth]{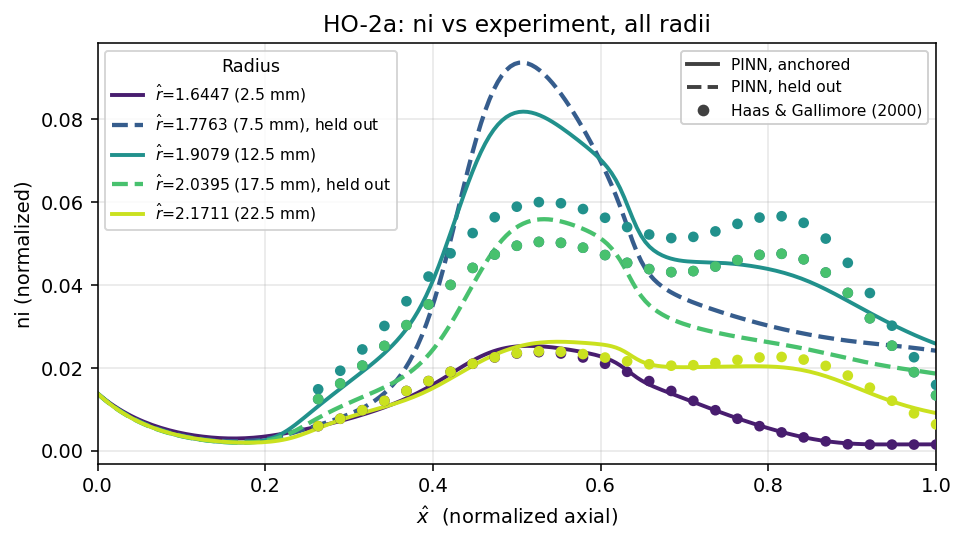}
\caption{HO-2a: 7.5 and \SI{17.5}{\milli\meter} withdrawn.}
\end{subfigure}\\[4pt]
\begin{subfigure}[t]{0.49\textwidth}
\includegraphics[width=\textwidth]{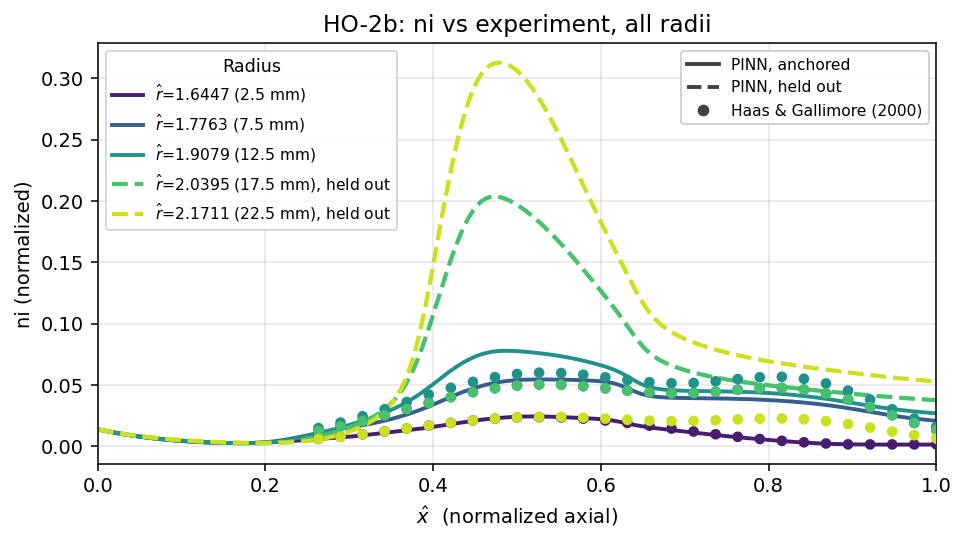}
\caption{HO-2b: 17.5 and \SI{22.5}{\milli\meter} withdrawn.}
\end{subfigure}
\caption{Ion density predicted by the three held-out runs against the reconstructed Haas--Gallimore profiles at all five radii. Solid lines: anchored radii; dashed lines: withdrawn radii. Note the different vertical scales.}
\label{fig:ho_ni}
\end{figure}

\begin{figure}[ht]
\centering
\begin{subfigure}[t]{0.32\textwidth}
\includegraphics[width=\textwidth]{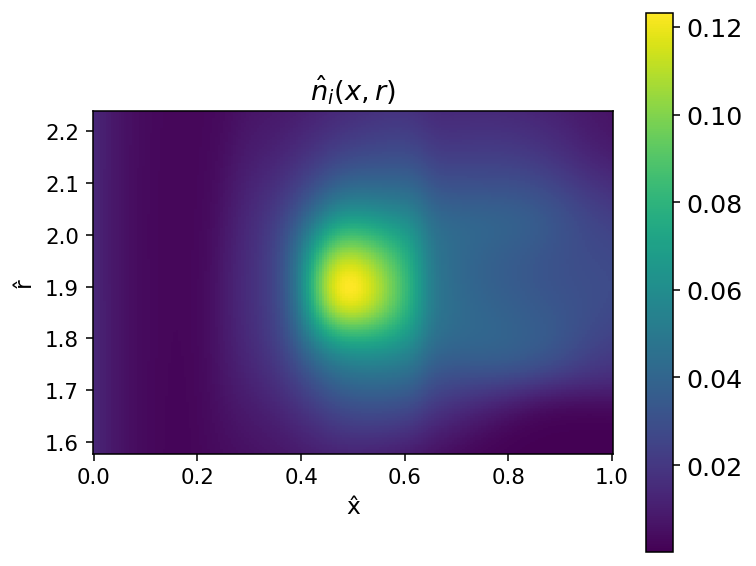}
\caption{HO-1.}
\end{subfigure}\hfill
\begin{subfigure}[t]{0.32\textwidth}
\includegraphics[width=\textwidth]{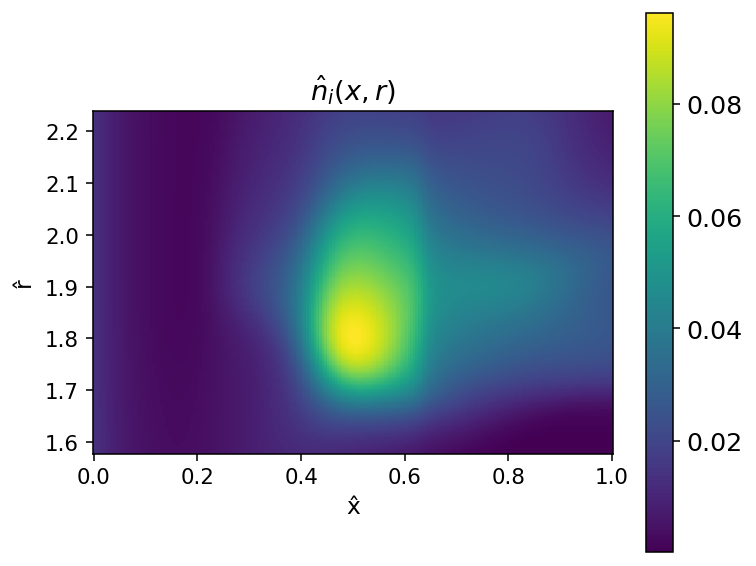}
\caption{HO-2a.}
\end{subfigure}\hfill
\begin{subfigure}[t]{0.32\textwidth}
\includegraphics[width=\textwidth]{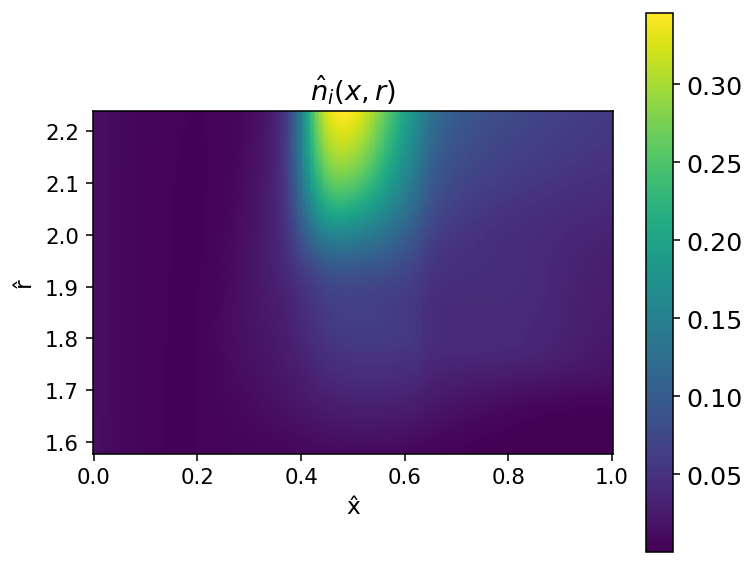}
\caption{HO-2b.}
\end{subfigure}
\caption{Converged ion-density maps of the three held-out runs. Compare with the fully anchored solution of Fig.~\ref{fig:fields}b.}
\label{fig:ho_maps}
\end{figure}

The composite loss does not reveal which fields the physics supplies and which the data supply, since both contribute to every field at once. To separate them, the model is retrained with some of the radial measurement lines removed from the objective and then evaluated at all five. The $n_i$, $T_e$, and $V_p$ anchors on a withdrawn line are removed together, the data weights are unchanged at $(10,5,30)$, and the configuration is otherwise identical to the reference run. Three cases are reported. HO-1 withdraws the central line at \SI{12.5}{\milli\meter} from the inner wall; HO-2a withdraws the two lines on either side of it, at 7.5 and \SI{17.5}{\milli\meter}; and HO-2b withdraws the two outer lines, at 17.5 and \SI{22.5}{\milli\meter}. Table~\ref{tab:heldout} gives the errors, Fig.~\ref{fig:ho_ni} the ion density at all five radii, and Fig.~\ref{fig:ho_maps} the converged ion-density maps.

\begin{table}[ht]
\centering
\caption{Held-out tests. Relative $L^2$ error of the ion density at the withdrawn radii, relative $L^2$ error of the potential over all five radii, predicted radial spread of the ion density at mid-channel relative to the measured spread ($0.0354$ normalized), and maximum radial gradient of the potential. The fully anchored reference run is listed for comparison.}
\label{tab:heldout}
\setlength{\tabcolsep}{5pt}
\begin{tabular}{llcccc}
\toprule
Case & Withdrawn & $n_i$ $L^2$ & $V_p$ $L^2$ & $n_i$ spread & $\max|\partial_r\phi|$ \\
 & (from inner wall) & (withdrawn) & (all five) & ratio & \\
\midrule
Reference & --- & --- & 0.01732 & --- & --- \\
HO-1 & \SI{12.5}{\milli\meter} & 0.587 & 0.01734 & 2.78 & 0.228 \\
HO-2a & 7.5, \SI{17.5}{\milli\meter} & 0.406 & 0.01733 & 1.93 & 0.131 \\
HO-2b & 17.5, \SI{22.5}{\milli\meter} & 3.444 & 0.01774 & 7.96 & 0.939 \\
\bottomrule
\end{tabular}
\end{table}

The potential error over all five radii is almost unaffected in every case, holding between $0.01733$ and $0.01774$ against $0.01732$ for the fully anchored reference, even where the withdrawn-radius density error is large. That is consistent with a field constrained by an elliptic equation anchored at both axial boundaries, although the experiment does not isolate that cause from the boundary values, the remaining anchors and the bounded output parametrization; a control without Poisson's equation would be needed for that attribution. The qualification is that the exit potential imposed as a boundary condition is itself taken from the reconstructed dataset.

The ion density behaves differently. It does not lose radial structure at the withdrawn radii; it produces too much of it. The predicted radial spread at mid-channel is 2.78 times the measured spread for HO-1, 1.93 for HO-2a, and 7.96 for HO-2b. Where the anchors are absent the solution is not flattened but unconstrained, and the ordering of the three cases follows from that. HO-2a withdraws twice as much information as HO-1 and reconstructs the density better, $0.406$ against $0.587$. The explanation is geometric rather than statistical: the measured ion density peaks at mid-channel, on the line HO-1 removes and HO-2a retains, and without it the HO-1 prediction overshoots the central density by roughly a factor of two (Fig.~\ref{fig:ho_ni}a). What governs radial reconstruction is whether the curvature of the profile is anchored, not how many lines are.

HO-2b fails, and fails differently. Its error of $3.444$ is not a degraded reconstruction but a divergence: at mid-channel the predicted density at the two withdrawn radii reaches about $0.20$ and $0.31$ in normalized units, against measured values of about $0.05$ and $0.025$ (Fig.~\ref{fig:ho_ni}c), and the largest radial gradient of the potential reaches $0.939$ against $0.228$ for HO-1 and $0.131$ for HO-2a, localized in the unanchored outer region. Withdrawing the two outermost lines leaves that region with no constraint from either the data or a boundary condition, and the solution there is unconstrained rather than inferred. Its total loss of $6.81$ is the lowest of the three, an artifact of where its anchors sit rather than how many it carries: HO-2a withdraws the same number of lines and converges to a higher loss and a better solution, a further instance of the aggregate loss being a poor guide to physical accuracy.

Taken together, the three cases separate the fields clearly. The potential remains substantially more robust than the ion density under the tested withdrawals, whereas the radial structure of the ion density remains strongly dependent on the measurement anchors. The electron temperature sits between the two. In HO-1 and HO-2a it is recovered at the withdrawn radii near and downstream of its peak, whereas in HO-2b it departs at the withdrawn outer radii, most strongly upstream, where it falls to about \SI{6}{\electronvolt} against a measured \SI{16}{\electronvolt} at $\hat{x}=0.3$. At that station the prediction sits below the data at every radius, anchored or not, in all three cases, by as little as 0.2 eV near the walls and up to about 7 eV on the central line, so the upstream deficit is a property of the model rather than of the withdrawal. Section~\ref{sec:bohm} shows that the electron transport parameters also affect the solution in ways the measured errors do not reveal, so how much radial variation the density can carry is not settled by the anchoring alone.

\FloatBarrier

\section{Physical-Consistency Analysis and Model Refinements}
\label{sec:consistency}

The reference solution of Sec.~\ref{sec:results} was judged by its fit to the reconstructed profiles and by the magnitude of its local residuals. This section tests it instead against integral and physical consistency requirements, and addresses the limitations it exposes by successive groups of corrections; several stages change more than one component at once, so the individual effect of each modification is not isolated. Every run is warm-started from the converged checkpoint of the preceding one, trained with the optimizer settings of Sec.~\ref{sec:arch}, and evaluated on the same grid; Table~\ref{tab:sequence} summarizes the sequence.

\subsection{Consistency diagnostics}
\label{subsec:diag}

Four diagnostics are evaluated on converged checkpoints without further training. The \emph{global balances} integrate the ion, neutral and electron fluxes over the channel cross-section at 40 axial stations, together with the wall fluxes and the ionization source between them; the ion imbalance is the difference between the ions created and those leaving through the exit and the walls, divided by the ions created, and the neutral imbalance is defined against the anode inflow. The network discharge current $I_d(x) = 2\pi e\int(n_iV_{ix} - n_eV_{ex})\,r\,dr$ is evaluated at the same stations.

The \emph{momentum-implied current} is obtained by neglecting electron inertia and solving the electron momentum equation locally for the electron velocity given the potential, the density, the temperature and the magnetic field. With $\mathbf{B} = (B_x, B_r, 0)$ the $3\times3$ linear system is
\begin{equation}
m_e\nu\,\mathbf{V}_e^{DD} = -e\left(\mathbf{E} + \mathbf{V}_e^{DD}\times\mathbf{B}\right) - \frac{\nabla(n_eT_e)}{n_e},
\label{eq:dd}
\end{equation}
with $\nu$ the total momentum-transfer frequency including the Bohm term. Integrating the axial electron flux gives a drift--diffusion discharge current $I_{DD}(x)$, which is compared with $I_d(x)$, and decomposing $\mathbf{V}_e^{DD}$ along and across $\mathbf{b} = \mathbf{B}/|\mathbf{B}|$ separates the field-aligned contribution from the cross-field and Hall contribution.

The \emph{field-aligned force imbalance} follows from projecting the electron momentum equation along $\mathbf{b}$ and neglecting inertia and parallel drag, which gives the equilibrium $\mathbf{b}\cdot\nabla\phi - \mathbf{b}\cdot\nabla(n_eT_e)/n_e \approx 0$ with $T_e$ in electron volts. Its departure from zero,
\begin{equation}
R_\parallel = \mathbf{b}\cdot\nabla\phi - \frac{\mathbf{b}\cdot\nabla(n_eT_e)}{n_e},
\label{eq:rpar}
\end{equation}
is reported in \si{\volt\per\meter}, as a cross-sectional RMS at each station and as an RMS over the domain. Finally, the \emph{ion momentum budget} integrates the terms of Eq.~\eqref{eq:ionmomx} over the cross-section at each station: the convective acceleration, the electric force, the ion--neutral exchange and the ionization birth momentum, whose sum is the residual.

\subsection{Global balances and the anode condition}
\label{subsec:global}

Applied to the reference solution, the global balances do not close. The anode neutral density corresponds to an inflow 6.1 times the measured flow of \SI{63}{sccm}, equivalent to \SI{27.5}{\ampere} against \SI{4.52}{\ampere}. The ionization source creates the equivalent of \SI{10.0}{\ampere} of ions, of which \SI{5.96}{\ampere} leave through the exit and a negligible amount through the walls, so that 41\% of the ions created do not leave the domain; since the model contains no sink for them, this is a global violation of ion continuity accumulated from local residuals that are each small. The exit ion current also exceeds the discharge current of \SI{5.39}{\ampere}.

The first correction set the anode neutral density from the measured flow, $n_{n,a} = \SI{5.50e18}{\per\meter\cubed}$, replaced the constant ion--neutral rate by a local charge-exchange rate computed from the neutral density and the relative ion--neutral speed, and added the global ion and neutral balances as integrated constraints. At the same time the electron axial velocity was no longer obtained from the closure of Eq.~\eqref{eq:closure} but emitted by the network and constrained by the electron momentum equations. Introduced incrementally from the reference checkpoint, these constraints were satisfied to $-1.6$ and $-2.0$\% for ions and neutrals; introduced together from a random initialization, they did not converge.

The reference configuration also fixes the anode potential to the \SI{300}{\volt} discharge voltage, whereas the reconstructed potential shows a plateau of \SI{274.04}{\volt} near the anode; the difference corresponds to an anode fall that the model does not resolve. With the \SI{300}{\volt} condition and the Gaussian field, which falls to \SI{3.9e-5}{\tesla} at the anode, the electrons there are essentially unmagnetized ($\omega_{ce}/\nu \approx 1.1$), and the potential gradient implies $I_{DD} = \SI{48789}{\ampere}$ at $\hat{x} = 0$. Setting the anode condition to \SI{274.04}{\volt}, with all else unchanged, reduced $I_{DD}$ to \SI{1371}{\ampere} at $\hat{x} = 0$ but only from 10{,}907 to \SI{2640}{\ampere} at $\hat{x} = 0.05$ and from 222 to \SI{158}{\ampere} at $\hat{x} = 0.2$: the condition alone does not make the near-anode region consistent with the electron momentum equation. The \SI{274.04}{\volt} condition is retained in every subsequent run, as is the \SI{116.2}{\volt} exit condition.

\subsection{Measured divergence-free magnetic field}
\label{subsec:bfield}

\begin{figure}[ht]
\centering
\includegraphics[width=\textwidth]{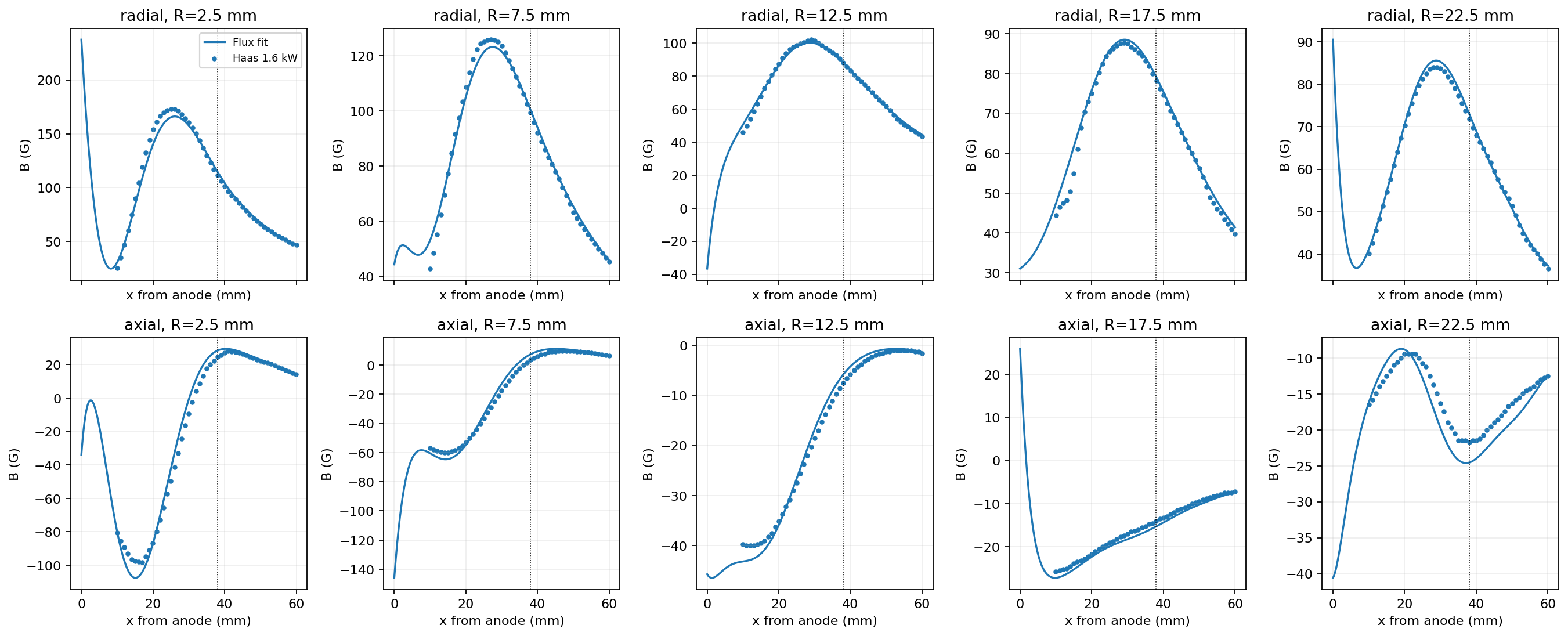}
\caption{Flux-function fit of the vacuum magnetic field (lines) against the Hall-probe measurements of Haas \citep{haas2001thesis} at \SI{1.6}{\kilo\watt} (symbols), for the radial (top) and axial (bottom) components at the five measured radii.}
\label{fig:bfit}
\end{figure}

The Gaussian field of Eq.~\eqref{eq:Bfield} has no axial component and no divergence-free structure. It was replaced by a two-dimensional field fitted to the vacuum Hall-probe measurements of Haas \citep{haas2001thesis} at the \SI{1.6}{\kilo\watt} operating point, for both components at five radial positions between 10 and \SI{60}{\milli\meter} from the anode, extracted directly from the vector graphics of the dissertation (51 points per curve, axis-calibration residual below \SI{0.05}{G}). The field is represented by a flux function $\psi(x,r)$,
\begin{equation}
B_x = \frac{1}{r}\frac{\partial\psi}{\partial r}, \qquad B_r = -\frac{1}{r}\frac{\partial\psi}{\partial x},
\label{eq:flux}
\end{equation}
so that $\nabla\cdot\mathbf{B} = 0$ holds by construction, and the fit minimizes the data misfit together with a penalty on the curl, the vacuum condition that constrains the extrapolation upstream of \SI{10}{\milli\meter}, where no measurements exist. The RMS errors are \SI{2.96}{G} for $B_x$ and \SI{3.09}{G} for $B_r$, within the probe-position uncertainty of $\pm\SI{1}{\milli\meter}$ given gradients up to \SI{13}{G\per\milli\meter} near the inner wall (Fig.~\ref{fig:bfit}). The electron momentum equations were extended to both field components, the Lorentz force now carrying $V_{e\theta}B_x$ in the radial component and $V_{ex}B_r - V_{er}B_x$ in the azimuthal one, and $|\mathbf{B}|$ entering the cyclotron and Bohm frequencies.

\begin{figure}[ht]
\centering
\includegraphics[width=\textwidth]{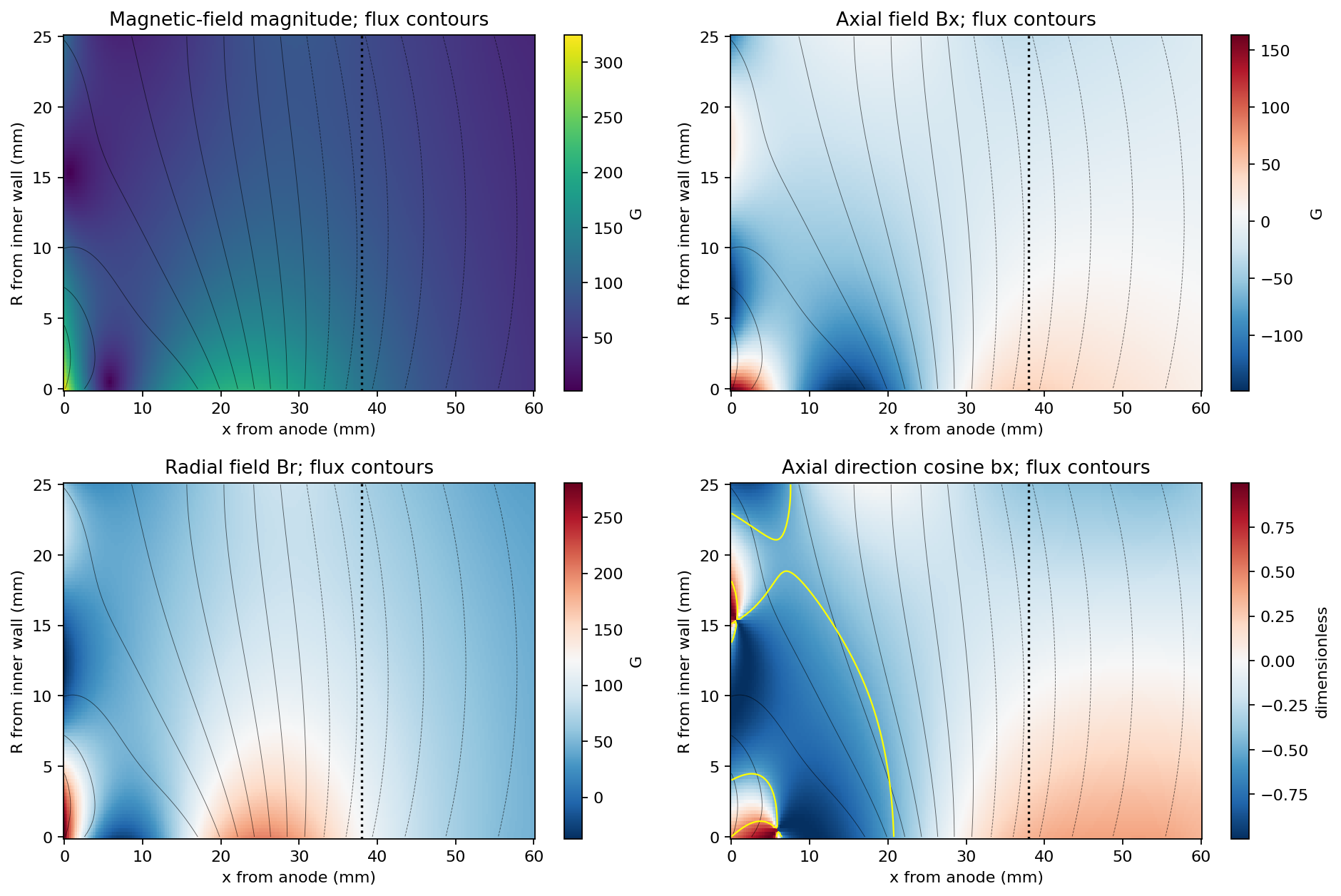}
\caption{Fitted vacuum magnetic field over the channel and near plume: magnitude, axial and radial components, and axial direction cosine $b_x = B_x/|\mathbf{B}|$, with flux contours (field lines). The dotted line marks the exit plane; the yellow contours in the last panel mark $|b_x| = 0.5$.}
\label{fig:bmaps}
\end{figure}

The measured field differs substantially from the Gaussian. At mid-radius it peaks at \SI{102}{G} at \SI{29}{\milli\meter}, where the Gaussian assumes \SI{200}{G}; at \SI{10}{\milli\meter} it is about \SI{61}{G} against \SI{14}{G}; and the axial component reaches $-40$ to $\SI{-80}{G}$ upstream, so the field lines are strongly inclined. Within \SI{2}{\milli\meter} of the anode and the inner wall the curl-free continuation rises to \SI{325}{G}, which is not supported by data; it affects about 1.3\% of the domain, and two fits with different curl weights agree to within \SI{8}{G} from \SI{5}{\milli\meter} onward, so results at $\hat{x} = 0$ are read as extrapolation-dependent. Figure~\ref{fig:bmaps} shows the fitted field and its field lines.

\begin{figure}[ht]
\centering
\includegraphics[width=\textwidth]{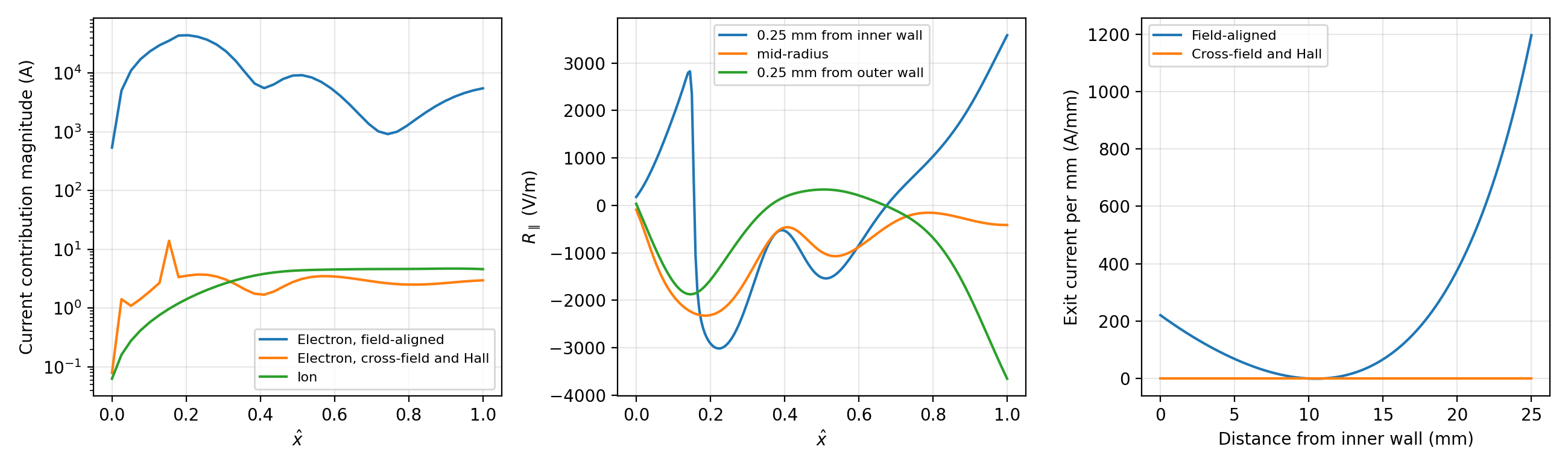}
\caption{Decomposition of the momentum-implied current with the measured field. Left: magnitudes of the field-aligned electron, cross-field and Hall electron, and ion contributions along the channel. Centre: field-aligned force imbalance $R_\parallel$ near each wall and at mid-radius. Right: radial distribution of the exit current.}
\label{fig:split}
\end{figure}

With the measured field, the cross-field and Hall part of the momentum-implied electron current lies between 0.1 and \SI{3.8}{\ampere} everywhere in the channel, the same order as the discharge current, whereas more than 99\% of the implied current is field-aligned (Fig.~\ref{fig:split}). The field-aligned force imbalance has an RMS of about \SI{1200}{\volt\per\meter}. Magnetic confinement acts across the field and not along it, so a force imbalance that is small in the normalized residual produces a large parallel electron velocity. The network current became positive along the channel, 1.3 to \SI{4.2}{\ampere}, but the implied current remained between about \SI{80}{\ampere}, at the field maximum, and \SI{4e4}{\ampere}.

\subsection{Bound of the potential representation}
\label{subsec:phibound}

\begin{figure}[ht]
\centering
\includegraphics[width=0.6\textwidth]{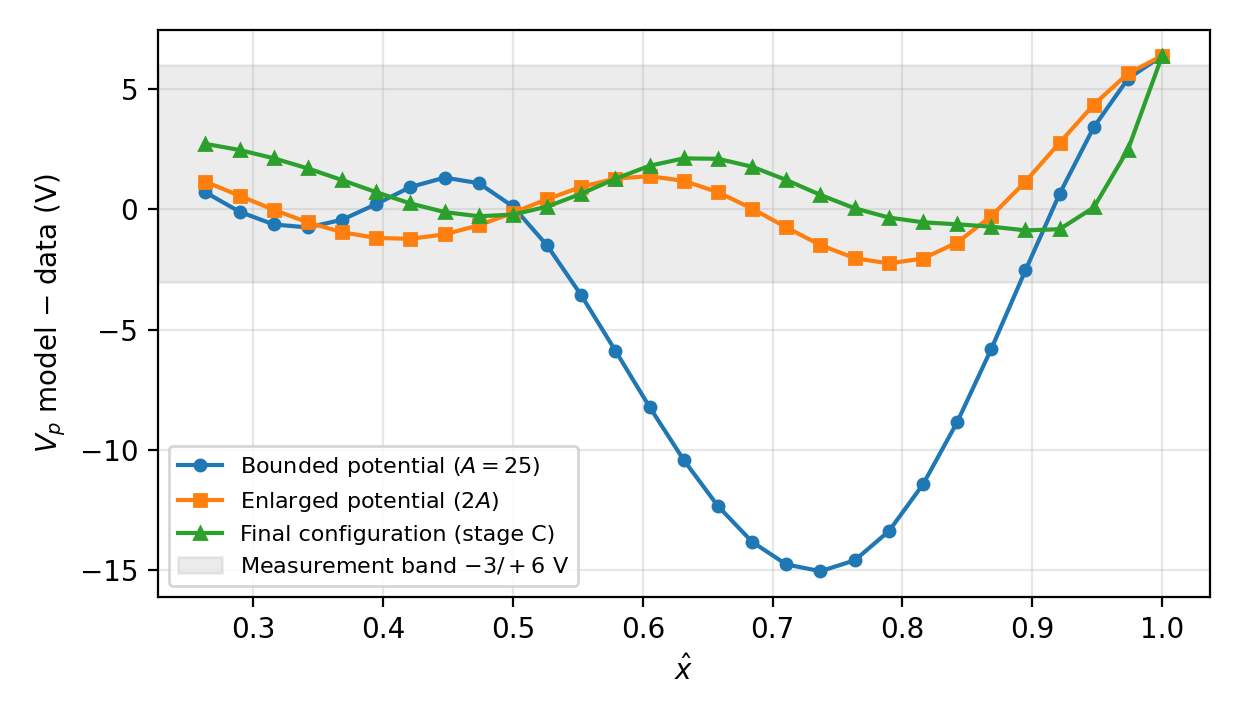}
\caption{Difference between the predicted and reconstructed plasma potential, averaged over the five radii, with the bounded representation, the enlarged representation, and the final configuration of Sec.~\ref{subsec:stageB}. The shaded region is the \SIrange{-3}{+6}{\volt} measurement uncertainty.}
\label{fig:vperr}
\end{figure}

Neither the anode condition nor the measured field changed the potential error in the acceleration region, which remained at \SI{-15}{\volt} at $\hat{x} = 0.74$. The cause is the representation of Eq.~\eqref{eq:phi_param}, which bounds the correction to the linear profile by $\xi(1-\xi)A_\phi\phi_{\mathrm{ref}}$, or $\xi(1-\xi)\times\SI{302}{\volt}$. Evaluating the correction that the reconstructed potential requires at each measurement point shows that 75 of the 145 points lie beyond this bound with the anode at \SI{274.04}{\volt}, by up to 27\%; with the \SI{300}{\volt} anode of the reference configuration, 50 points already do, which is the origin of the local potential deficit noted in Sec.~\ref{sec:results}. In the converged solution $|\tanh(\hat{u}_0)| > 0.95$ at 101 of the 145 points, and the saturation coincides with the region where the potential departs from the measurements. The bound was doubled by writing the correction as $2A_\phi\tanh(\hat{u}_0/2)$, which equals the original form to first order and therefore preserves the warm start wherever the network was not saturated. With this change alone, no point is saturated, the relative $L^2$ error of the potential falls from 0.033 to 0.009, and the potential lies within the \SIrange{-3}{+6}{\volt} band everywhere except at the exit plane, where it is set by the \SI{116.2}{\volt} condition against a reconstructed \SI{109.8}{\volt} (Fig.~\ref{fig:vperr}). Particle conservation is unaffected, at $-1.1$ and $-1.8$\% for ions and neutrals.

\subsection{Ion momentum and the density--potential tension}
\label{subsec:ions}

\begin{figure}[ht]
\centering
\includegraphics[width=\textwidth]{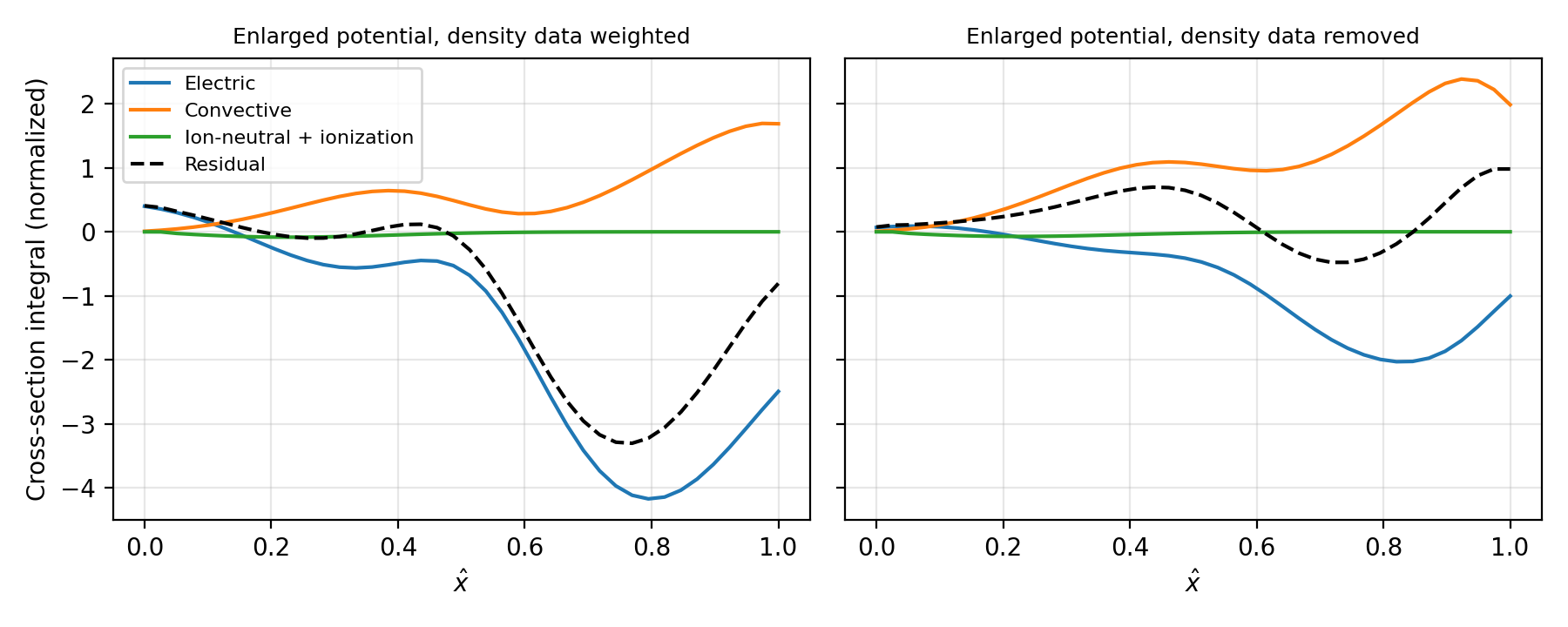}
\caption{Cross-section integral of the terms of the axial ion momentum residual along the channel, with the enlarged potential representation: (a) density measurements in the objective; (b) density measurements removed.}
\label{fig:ionbudget}
\end{figure}

Once the potential matches the measurements, the axial ion momentum budget shows that the residual is concentrated between $\hat{x} = 0.5$ and 0.9 and is an imbalance between two terms only: the electric force exceeds the convective acceleration by a factor of about six, $-3.7$ against $+0.6$ at $\hat{x} = 0.72$ (Fig.~\ref{fig:ionbudget}a), while the ion--neutral exchange and the ionization birth momentum are negligible there. The ions do not accelerate as much as the measured potential drop requires and leave at \SI{9.2}{\kilo\meter\per\second}. A ballistic estimate confirms the size of the discrepancy: ions born at the locations of the model ionization source, 93\% of which lies upstream of $\hat{x} = 0.5$, and falling collisionlessly through the reconstructed potential reach a mean exit velocity of \SI{15.5}{\kilo\meter\per\second}. The reconstructed density, through the measured mass flow, limits the exit velocity to about \SI{9}{\kilo\meter\per\second}, and laser-induced fluorescence gives \SI{11}{\kilo\meter\per\second} \citep{williams1999}. Removing the density measurements from the objective, with everything else unchanged, reduced the ion momentum residual from 0.067 to 0.0054 and raised the exit velocity to \SI{16.7}{\kilo\meter\per\second}, with 97\% of the potential points within their band but only 31\% of the density points within $\pm50\%$ (Fig.~\ref{fig:ionbudget}b).

Within the tested reduced single-velocity fluid formulation and its current closures, the reconstructed density, the measured potential and the measured ion velocity could therefore not be matched simultaneously; this is a statement about the formulation, which omits ion pressure and prescribes its closures, not about every single-velocity description. Ionization extending into the acceleration region, a spread of ion birth potentials, or probe perturbation of the potential measurement are possible explanations; they are not tested here, and an analysis of the ion birth location and velocity distribution is required before the inconsistency is attributed to the measurements.

\subsection{Uncertainty-weighted data and electron pressure}
\label{subsec:stageA}

Rather than removing a measurement, the data terms were reformulated so that each residual is normalized by its measurement uncertainty,
\begin{equation}
z_{n} = \frac{n_i - n_i^{\mathrm{obs}}}{0.5\,n_i^{\mathrm{obs}}}, \qquad
z_{T} = \frac{T_e - T_e^{\mathrm{obs}}}{0.2\,T_e^{\mathrm{obs}}}, \qquad
z_{V} = \frac{V_p - V_p^{\mathrm{obs}}}{\sigma_V}, \quad
\sigma_V = \begin{cases}\SI{6}{\volt} & V_p > V_p^{\mathrm{obs}},\\ \SI{3}{\volt} & \text{otherwise},\end{cases}
\label{eq:zscores}
\end{equation}
each data term being the mean of $z^2$ with unit weight, so that an error equal to the measurement uncertainty costs one in every quantity. At the same time the electron pressure gradient was added to the axial and radial electron momentum components, with unit normalized coefficient since $\phi_{\mathrm{ref}} = T^{*}$; it must be present before the field-aligned balance it governs can be imposed. The converged result (stage A) places 96\% of the potential points within \SIrange{-3}{+6}{\volt}, 90\% of the temperature points within $\pm20\%$ and 38\% of the density points within $\pm50\%$, with ion and neutral imbalances of $-0.7$ and $+3.6$\% and an exit velocity of \SI{15.8}{\kilo\meter\per\second}. Weighted by its stated uncertainty, the density remains the measurement the solution departs from, and the network current is \SI{-0.7}{\ampere} at the anode and \SI{2.9}{\ampere} at the exit.

\subsection{Field-aligned constraint and current condition}
\label{subsec:stageB}

\begin{figure}[ht]
\centering
\begin{subfigure}[t]{0.49\textwidth}
\includegraphics[width=\textwidth]{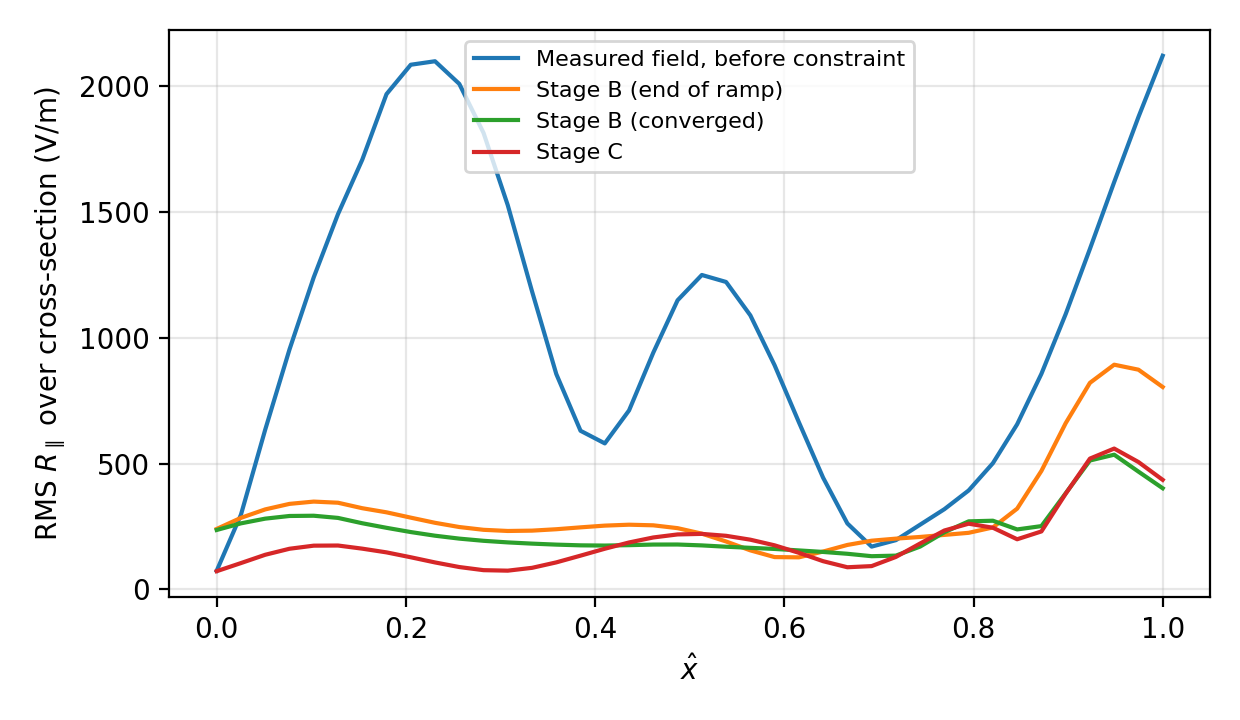}
\caption{RMS of $R_\parallel$ over the cross-section.}
\label{fig:rpar}
\end{subfigure}\hfill
\begin{subfigure}[t]{0.49\textwidth}
\includegraphics[width=\textwidth]{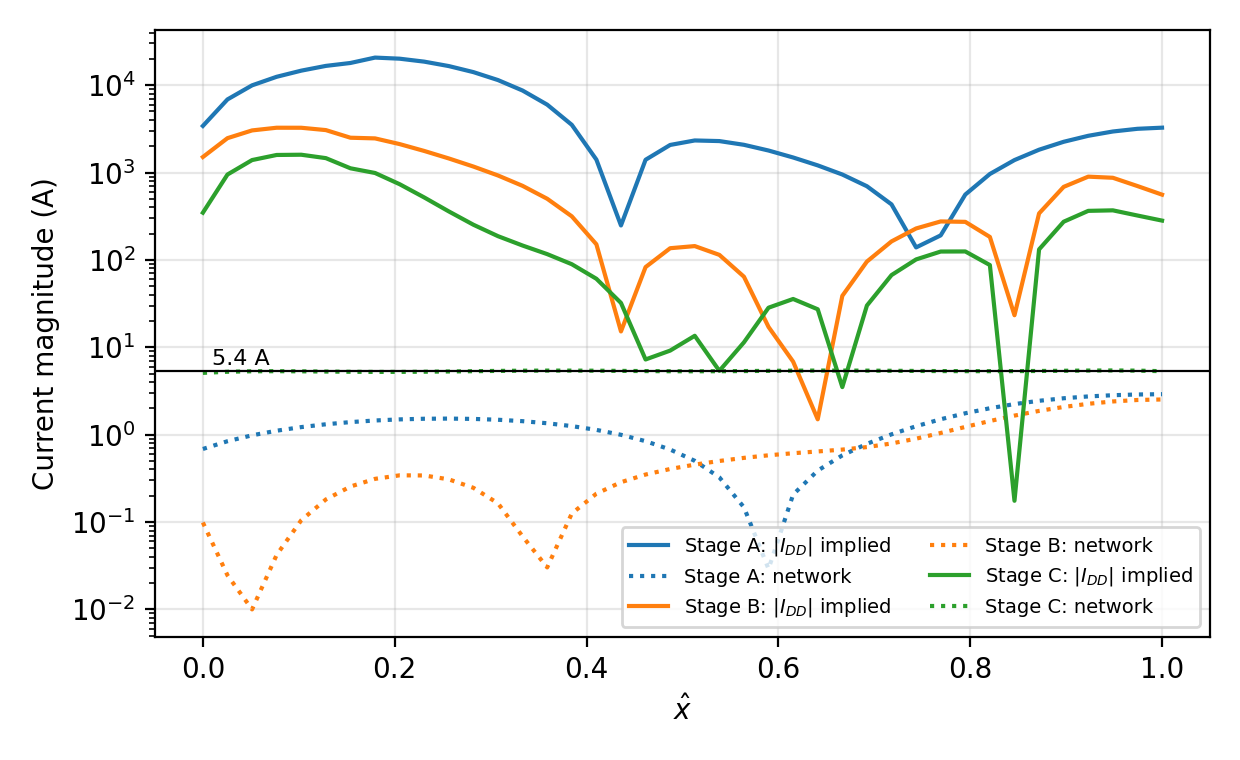}
\caption{Implied $|I_{DD}|$ (solid) and network current (dotted).}
\label{fig:currents}
\end{subfigure}
\caption{Field-aligned constraint and current condition. (a) Cross-sectional RMS of the field-aligned force imbalance $R_\parallel$ along the channel, with the measured field before the constraint, at the end of the ramp, converged (stage B), and with the current condition added (stage C). (b) Magnitude of the momentum-implied discharge current $|I_{DD}|$ and of the network discharge current for stages A, B and C; the horizontal line marks the measured \SI{5.4}{\ampere}.}
\label{fig:fieldaligned}
\end{figure}

A direct penalty on $\mathbf{V}_e - \mathbf{V}_e^{DD}$ is poorly conditioned: its initial value was about $7\times10^{8}$, and the optimizer sacrificed the density fit and ion conservation to reduce it. The field-aligned balance was instead imposed through the projection of the complete retained electron momentum equation, including inertia and drag so that a non-zero parallel current is not excluded, normalized by a reference field,
\begin{equation}
\mathcal{L}_\parallel = \left\langle\left(\frac{\mathbf{b}\cdot\mathbf{R}_e}{n_e\,E_{\mathrm{ref}}}\right)^2\right\rangle, \qquad E_{\mathrm{ref}} = \SI{1200}{\volt\per\meter},
\label{eq:lpar}
\end{equation}
where $\mathbf{R}_e$ is the vector of electron momentum residuals expressed in \si{\volt\per\meter}. Its weight was raised from zero to ten along $\lambda(s) = \lambda_{\mathrm{target}}[1-\cos(\pi s)]/2$ over 3{,}000 epochs, with the learning-rate schedule suspended during the ramp, since the total loss rises by construction and would otherwise drive the rate to its floor; the run was then continued at full weight for a further 8{,}000 epochs (stage B). The constraint reduced the RMS of $R_\parallel$ from about 950 to \SI{257}{\volt\per\meter} (Fig.~\ref{fig:rpar}) and the implied current by one to two orders of magnitude, while particle conservation, the data agreement and the ion momentum residual remained stable. It did not, however, determine the discharge current, which oscillated between $-1.7$ and \SI{4.9}{\ampere} during training and converged between $-0.1$ and \SI{2.5}{\ampere}. The integrated current condition $I_d(x) = \SI{5.4}{\ampere}$ at the 40 stations was therefore added, ramped in the same way to a weight of 100 over 2{,}000 epochs, with the field-aligned weight held at ten (stage C).

With both constraints, the network current remains between 5.1 and \SI{5.4}{\ampere} along the channel under the imposed integral-current penalty, the RMS of $R_\parallel$ falls further to \SI{230}{\volt\per\meter}, 97\% of the potential points and 88\% of the temperature points lie within their bands, and the ion and neutral imbalances are $-2.7$ and $-2.3$\%. Between $\hat{x} = 0.45$ and 0.85 the implied current falls to between about 4 and \SI{90}{\ampere}, but it remains of order $3\times10^{2}$ to \SI{1.6e3}{\ampere} near the anode and about \SI{280}{\ampere} at the exit (Fig.~\ref{fig:currents}). Since the implied parallel current scales directly with $R_\parallel$, agreement at the level of one ampere near the anode would require $R_\parallel$ of order 0.2 to \SI{1}{\volt\per\meter}, about three orders of magnitude below the normalization scale. The cost of the current condition falls on the ions: the exit velocity rises to \SI{21.8}{\kilo\meter\per\second}, twice the fluorescence value, only 10\% of the density points remain within their band, and the exit ion current is 0.55 of the discharge current against a measured beam fraction of about 0.76.

\subsection{Exit ion flux and collisionless ion birth distribution}
\label{subsec:birth}

The exit ion current and velocity of stage C were evaluated directly by cross-sectional integration, $I_i = 2\pi e\int n_iV_{ix}\,r\,dr$ and $\overline{V}_{ix,n} = \int n_iV_{ix}r\,dr/\int n_ir\,dr$. The exit ion current is \SI{2.93}{\ampere}, 28\% below the \SI{4.10}{\ampere} implied by the measured beam fraction of 0.76, and the density-weighted exit velocity is \SI{21.1}{\kilo\meter\per\second}, close to the centerline value, so the excess velocity is not a centerline artifact. The integrated ionization source equals \SI{2.94}{\ampere} of ions, a propellant utilization of 65\%, whereas the measured beam current implies about 91\%. With the current condition imposed, the network therefore ionizes too little propellant and accelerates the ions it creates too much: a kinetic energy of about \SI{300}{\electronvolt} at \SI{21.1}{\kilo\meter\per\second} exceeds the \SI{158}{\volt} potential drop available in the channel, so the stage C ion velocity is not consistent with energy conservation. Integrating the axial ion momentum equation along lines of constant radius at a quarter, half and three quarters of the channel width splits the kinetic energy gain of stage C, 346, 322 and \SI{294}{\electronvolt}, into \SI{158}{\electronvolt} of electric work, 237, 213 and \SI{232}{\electronvolt} of work done by the ion momentum residual, and the remaining physical terms (ionization loading, ion--neutral exchange, electron drag and radial convection, $-49$ to \SI{-96}{\electronvolt}); the residual RMS exceeds that of the electric force by a factor of 1.2 to 1.35. The excess velocity is therefore produced by the unsatisfied ion momentum equation: stage C improves the imposed integral constraints but does not provide an energetically consistent ion-acceleration solution, and its exit velocity is not by itself evidence about the limits of single-velocity ion fluids. In stage A, where the electric work accounts for 93 to 96\% of the kinetic energy gain at the same three radii, the exit ion current is \SI{4.38}{\ampere}, within 7\% of the measured estimate, the utilization is 96\%, and the density-weighted exit velocity of \SI{15.9}{\kilo\meter\per\second} matches the collisionless estimate below.

To test whether a distribution of ion birth locations could reconcile the measurements, ions were assumed to be created at the local ionization rate of the model, to start at the neutral velocity, and to move collisionlessly and axially to the exit, without radial motion, charge exchange or wall loss, each gaining $e[\phi(x',r)-\phi(x_{\mathrm{exit}},r)]$. With the potential and ionization fields of stage C, the resulting exit-velocity distribution has a flux-weighted mean of \SI{14.4}{\kilo\meter\per\second}, a density-weighted mean of \SI{13.7}{\kilo\meter\per\second}, and 10th to 90th percentiles of 12.3 and \SI{15.4}{\kilo\meter\per\second}; replacing the model potential by the reconstructed one gives 14.6, 14.1, 12.7 and \SI{15.5}{\kilo\meter\per\second}. Only 2\% of the ions leave below \SI{8}{\kilo\meter\per\second}. The distribution is therefore narrow, because most ions are created upstream of the main potential drop, and a single ion velocity is an adequate description of it. A mean velocity of \SI{11}{\kilo\meter\per\second} corresponds to a mean energy of about \SI{82}{\electronvolt}, which with the reconstructed potential requires the ions to be born on average near $\hat{x}\approx0.83$, inside the acceleration region, whereas 83\% of the model ionization lies upstream of $\hat{x}=0.7$. Within the collisionless assumptions, the fluorescence velocity and the reconstructed potential are reconciled only if ionization is concentrated well downstream of the electron temperature peak, or if the potential drop at the operating point of the fluorescence measurement lies further downstream than the probe data indicate.

\begin{table}[!htbp]
\centering
\caption{Sequence of consistency runs. Successive runs introduce groups of corrections, not always a single change; all runs after the anode correction use the \SI{274.04}{\volt} anode and \SI{116.2}{\volt} exit conditions, and from the fourth line the measured magnetic field. Band fractions are the share of measurement points within $\pm50\%$ ($n_i$), $\pm20\%$ ($T_e$) and \SIrange{-3}{+6}{\volt} ($V_p$); imbalances are for ions and neutrals; $I_d$ is the network discharge current at the anode and at the exit. For the reference, the ion imbalance is the share of created ions that do not leave the domain, and its neutral inflow is 6.1 times the measured flow (Sec.~\ref{subsec:global}).}
\label{tab:sequence}
\setlength{\tabcolsep}{5pt}
\begin{tabular}{lcccc}
\toprule
Run & Bands $n_i$/$T_e$/$V_p$ (\%) & Imbalance (\%) & $I_d$ anode/exit (A) & $V_{ix}$ exit (km/s) \\
\midrule
Reference (Sec.~\ref{sec:results}) & 98/91/66 & $-40$/-- & 5.39 (closure) & 5.5 \\
Measured inflow, balances & 43/74/63 & $-1.6$/$-2.0$ & $-5.2$/1.8 & 8.4 \\
Anode at \SI{274.04}{\volt} & 48/90/52 & $-0.8$/$-1.2$ & 0.1/3.5 & 9.0 \\
Measured magnetic field & 48/94/52 & $-0.3$/$-0.9$ & 1.4/4.2 & 9.4 \\
Enlarged potential & 46/82/97 & $-1.1$/$-1.8$ & $-1.0$/3.3 & 9.2 \\
Stage A: uncertainty, pressure & 38/90/96 & $-0.7$/$+3.6$ & $-0.7$/2.9 & 15.8 \\
Stage B: field-aligned & 25/87/95 & $-2.2$/$+0.7$ & $-0.1$/2.5 & 15.9 \\
Stage C: current & 10/88/97 & $-2.7$/$-2.3$ & 5.1/5.3 & 21.8 \\
\bottomrule
\end{tabular}
\end{table}

\FloatBarrier

\section{Conclusion}

We have presented, to our knowledge, the first two-dimensional axisymmetric physics-informed neural network that solves the coupled multifluid system of a Hall effect thruster discharge with self-consistent electrostatics, the potential being obtained from Poisson's equation together with the resolved charge densities through a bounded charge-separation field, and that assimilates sparse internal probe measurements of the P5 thruster directly into its objective. The reference solution reproduces the plasma potential, the electron temperature and the ion density with relative $L^2$ errors of 0.017, 0.073 and 0.205, and a published hybrid PIC-MCC simulation of the same thruster reproduces the potential and the density to a similar level while missing the electron temperature by a factor of two, the weakness the PINN reports as its largest equation residual. The held-out tests show that the potential survives the withdrawal of any one or two measurement lines, whereas the ion-density reconstruction depends on whether the radius at which the profile peaks remains anchored.

The physical-consistency analysis of Sec.~\ref{sec:consistency} shows that a good data fit is not evidence of a consistent solution, and three results stand independently of the questions that remain. First, the bounded potential representation of the reference configuration cannot reach part of the measured potential, which accounts for its potential deficit in the acceleration region; enlarging it reduces the potential error by a factor of almost four. Second, with a divergence-free field fitted to the measured vacuum field, the cross-field electron transport implied by the momentum equation is of the magnitude of the discharge current, and the remaining electron inconsistency is entirely field-aligned. Third, within the tested single-velocity fluid formulation and its current closures, the reconstructed ion density, the measured potential and the measured ion velocity could not be matched simultaneously: the final configuration conserves particles to within 3\%, holds the network current between 5.1 and \SI{5.4}{\ampere} and matches the potential and temperature measurements within their uncertainties, but it ionizes too little propellant and accelerates the ions beyond the available potential drop, the excess energy being supplied by the ion momentum residual. A collisionless birth-location estimate shows that the ion velocity distribution implied by the model ionization and the measured potential is narrow, centered near \SI{14}{\kilo\meter\per\second}, so that a broader distribution alone does not explain the \SI{11}{\kilo\meter\per\second} measured by fluorescence. The model is therefore best read as a physics-constrained reconstruction and diagnostic tool, whose reliability differs between fields and depends on the sensor layout, and which identifies which assumptions and which measurements prevent a self-consistent description.

\subsection{Limitations and future work}
\label{sec:limits}

The field-aligned electron balance must be tightened by roughly three orders of magnitude near the anode and the exit before the momentum-implied and network currents can agree there; a progressive reduction of $E_{\mathrm{ref}}$, each stage introduced only once the previous one is stable, is the direct way to attempt it. A collisional ion birth and transport calculation, including charge exchange and radial motion, should extend the collisionless estimate of Sec.~\ref{subsec:birth}; if it confirms a narrow distribution, the tension lies between the measurements themselves, and if not, a hybrid model retaining the electron fluid PINN with a kinetic or trajectory-based ion description is the natural extension. A frozen-heavy-species electron test with both the current condition and the field-aligned balance enforced would separate the electron closure from the ion description. The exit potential condition, \SI{6.4}{\volt} above the value reconstructed at the exit plane, should be tested in a separate controlled run, and the curl-free extrapolation of the magnetic field near the anode corner should be constrained by measurements or by a magnetostatic computation of the circuit. The energy equation carries no conductive flux and no wall losses, and its residual remains the largest of the reference solution. Finally, the weights of the composite loss are set by hand; adaptive balancing \citep{wang2022} would remove that step, and extending the inputs to the discharge voltage and mass flow would turn the single-point model into an operating-envelope surrogate. Internal Langmuir-probe maps \citep{linnell2006} and fluorescence velocity measurements \citep{macdonald2017} available for other thrusters would test whether the diagnostics reported here generalize beyond the P5.

\bibliographystyle{unsrtnat}
\bibliography{references}
\end{document}